\documentclass[
twocolumn,
amsmath,
amssymb,
prl,
superscriptaddress,  
nobibnotes,  
]{revtex4-2}

\usepackage{latexsym}
\usepackage{graphicx}
\usepackage{siunitx}
\usepackage{xcolor}
\usepackage{soul}
\usepackage{amsmath}
\usepackage{amssymb}
\usepackage{bm}
\usepackage{hyperref}
\usepackage[capitalise]{cleveref}
\DeclareUnicodeCharacter{03C0}{\ensuremath{\pi}}
\usepackage{upgreek}
\renewcommand{\pi}{\uppi}

\begin{document}

\title{Superconducting Diode Effect Sign Change in Epitaxial Al-InAs Josepshon Junctions}

\author{Neda Lotfizadeh}
\author{William F. Schiela}
\affiliation{Center for Quantum Information Physics, Department of Physics, New York University, New York, NY 10003, USA}

\author{Bar{\i}\c{s} Pekerten}
\affiliation{Department of Physics \& Astronomy, Wayne State University, Detroit, MI 48201, USA}
\affiliation{Department of Physics, University at Buffalo, State University of New York, Buffalo, NY 14260, USA}

\author{Peng Yu}
\author{Bassel Heiba Elfeky}
\author{William M. Strickland}
\affiliation{Center for Quantum Information Physics, Department of Physics, New York University, New York, NY 10003, USA}

\author{Alex Matos-Abiague}
\affiliation{Department of Physics \& Astronomy, Wayne State University, Detroit, MI 48201, USA}

\author{Javad Shabani}
\email[]{jshabani@nyu.edu}
\affiliation{Center for Quantum Information Physics, Department of Physics, New York University, New York, NY 10003, USA}

\date{\today}

\begin{abstract}\begin{center}\textbf{Abstract}\end{center}
There has recently been a surge of interest in studying the superconducting diode effect (SDE) partly due to the possibility of uncovering the intrinsic properties of a material system. A change of sign of the SDE at finite magnetic field has previously been attributed to different mechanisms.  Here, we observe the SDE in epitaxial Al-InAs Josephson junctions with strong Rashba spin-orbit coupling (SOC). We show that this effect strongly depends on the orientation of the in-plane magnetic field. In the presence of a strong magnetic field, we observe a change of sign in the SDE. Simulation and measurement of supercurrent suggest that depending on the superconducting widths, $W_\text{S}$, this sign change may not necessarily be related to 0--$\pi$ or topological transitions. We find that the strongest sign change in junctions with narrow $W_\text{S}$ is consistent with SOC-induced asymmetry of the critical current under magnetic-field inversion, while in wider $W_\text{S}$, the sign reversal could be related to 0--$\pi$ transitions and topological superconductivity.
\end{abstract}

\maketitle

\section{Introduction}

Nonreciprocity in non-centrosymmetric quantum systems has been well studied in semiconductors as they are essential for the rectification function in electrical diodes and solar cells. There has been a recent rise of interest in nonreciprocity in superconductors, implying a progress towards designing superconducting
diodes and its possible application in modern electronic circuits, sensors, and detectors \cite{ando_observation_2020, baumgartner_effect_2022, baumgartner_supercurrent_2022, bauriedl_supercurrent_2022, wu_field-free_2022, pal_josephson_2022, yuan_supercurrent_2022, mazur_gate-tunable_2022, Sundaresh2023:NC, gupta_superconducting_2022, zhang_evidence_2022,margineda2023}.  Nonreciprocal critical currents in superconductors occur when the magnitude of the critical supercurrent, $I_\text{c}$, depends on the direction in which the current is swept. Theoretically, the so-called diode effect can occur when both inversion and time reversal symmetries are broken, where the latter can be achieved by magnetic proximity effect, in magnetic Josephson junctions, or by applying an external magnetic field. This effect  has  been attributed to the presence of finite-momentum Cooper pairs and the change in the nature of superconductivity \cite{davydova_universal_2022, daido_intrinsic_2022, yuan_supercurrent_2022, ilic_theory_2022, he_phenomenological_2022}. Recent studies have suggested the existence of the superconducting diode effect (SDE) in Josephson junctions (JJs) with large Rashba spin-orbit coupling (SOC) \cite{baumgartner_effect_2022, Costa2022:arxiv, turini_josephson_2022, pal_josephson_2022, ando_observation_2020, wu_field-free_2022, jeon_zero-field_2022}. The magnitude of the supercurrent in JJs with SOC depends on the direction of the magnetic field, as the Rashba and Dresselhaus effects can have different contributions \cite{Pakizer2021:PRR,Pekerten2022:PRB}. Therefore, investigating the SDE through a JJ can provide information about the SOC in its semiconductor. 

Planar JJs fabricated on epitaxial Al-InAs heterostructures are great candidates to study SDE due to their strong SOC \cite{shabani_two-dimensional_2016, baumgartner_supercurrent_2022, baumgartner_effect_2022}. Such devices have also shown signatures of topological phase transition when their time reversal symmetry is broken by an in-plane magnetic field \cite{fornieri_evidence_2019, dartiailh_phase_2021, Banerjee2023:arxiv}. Recently, Costa \emph{et al.} \cite{Costa2022:arxiv} have reported a sign reversal of the AC SDE in multi-channel JJs based on Al-InAs with strong SOC subjected to a magnetic field, and related it to a 0--$\pi$-like transition induced by the Zeeman interaction in the device. Conversely, Banerjee \emph{et al.} \cite{Banerjee2023:arxiv} have proposed a SDE originating from finite-momentum Cooper pairing solely due to orbital effects, without invoking SOC or Zeeman interaction. 

In this work, we study epitaxial Al-InAs JJs with various superconductive contact widths, $W_{S}$. By applying a magnetic field perpendicular to the current and parallel to the junction, we observe nonreciprocal critical currents due to the finite-momentum Cooper pairing enabled by the coexistence of strong Rashba SOC and the Zeeman interaction. We observe a SOC-induced shift, $B_*$, of the magnetic field yielding the maximum of the critical current amplitude and use it to estimate the Rashba SOC strength in the JJ. In the absence of the magnetic field, time-reversal symmetry is restored and the SDE vanishes. However, the SDE can also vanish at certain finite magnetic fields and changes sign below the superconductor critical field, $B_\text{c}$. We consider JJs with various superconducting widths, $W_\text{S}$, and  observe zeros of the SDE, across which the critical current difference $\Delta I_\text{c}=I_\text{c}^+ - |I_\text{c}^-|$ characterizing the SDE exhibits sign reversals at finite values of the magnetic field. We attribute the sign reversals to i) 0--$\pi$-like jumps of the ground-state superconducting phase difference for wide $W_\text{S}$ and ii) SOC-induced asymmetry of the critical current under magnetic-field inversion with respect to the field-shift, $B_*$ for narrow $W_\text{S}$. In a gated junction, we observe the SDE and SOC-induced shift at zero and positive gate voltages where the SOC is strong in our system. However, the SDE is negligible when a negative gate voltage is applied, suggesting that the Rashba strength is relatively small at negative gate voltages. This agrees with our previous studies of SOC strength measurements on gated Hall bars.

\section{Results and Discussion}

\subsection{Devices and measurement details}

\begin{figure}[ht!]
    \centerline{\includegraphics[scale=0.55]{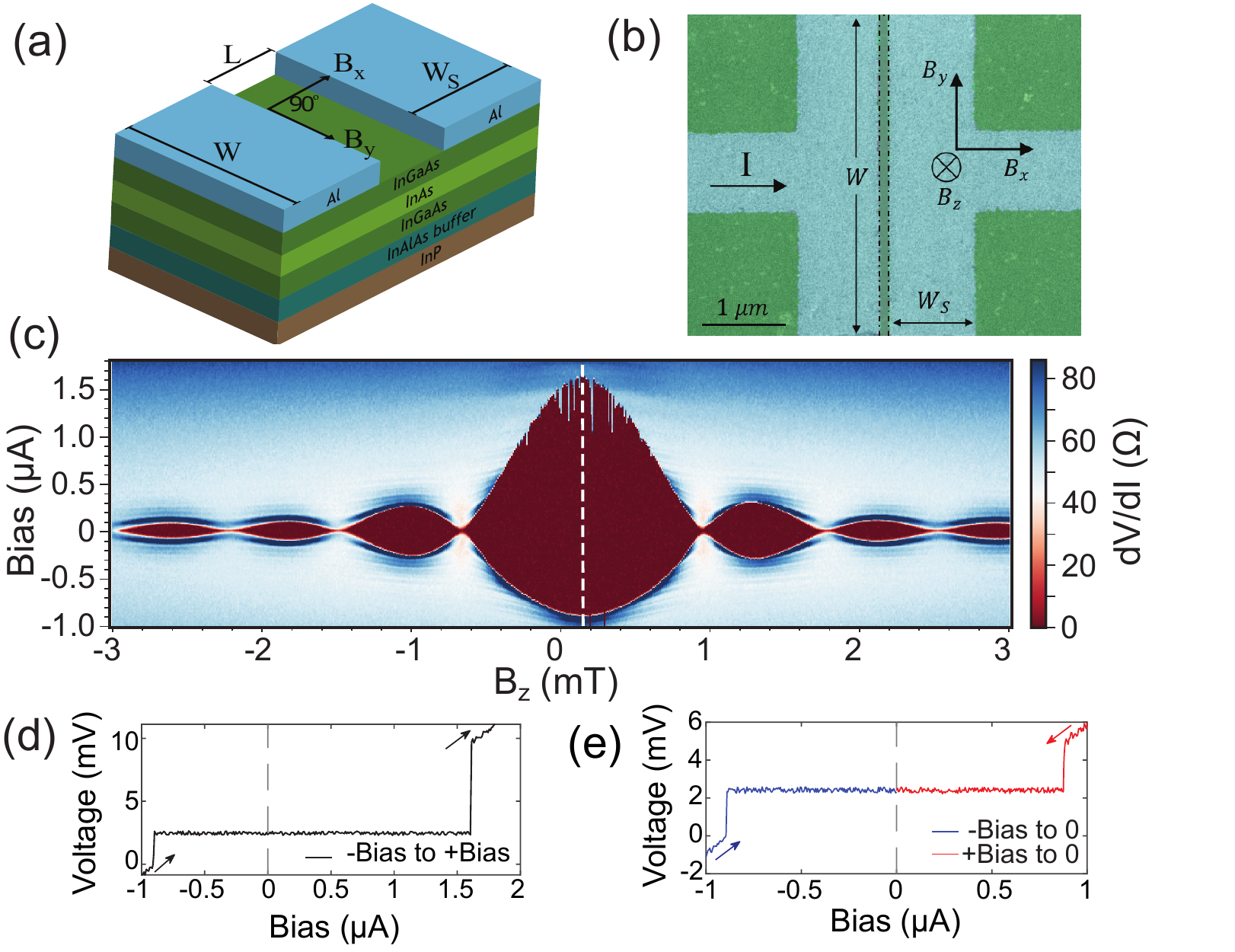}}
    \caption{\textbf{Devices and measurement methods.} (a) A schematic of a junction of length $L$, width $W$ and superconducting width of $W_\text{S}$ fabricated on the Al-InAs heterostructure. The superconducting contacts are made of Al and the quantum well (QW) consists of a layer of InAs grown between two layers of In$_{0.81}$Ga$_{0.19}$As. (b) False colored SEM image of a typical junction showing Al (blue) and QW (green) regions. The dashed line between the superconducting contacts is the $W=\SI{4}{\micro m}$ wide and $L=\SI{150}{\nano m}$ long etched gap.   The magnetic field can be applied in three direction independently as shown on the SEM image. (c) Differential resistance as a function of the bias current and out-of-plane magnetic field of Josephson junction 1 (JJ1)  with $W_\text{S}=$ $\SI{0.6}{\micro m}$  at zero in-plane magnetic field. A hysteresis due to the thermal effects can be seen. White dashed line indicates the position of the maximum of the critical current. (d) A line cut of (c) showing hysteresis in voltage  versus current when the bias is swept from negative to positive.  The values  of  supercurrent on each side are different due to  the thermal effects.  (e) Voltage  versus current when the bias is swept from negative to zero (blue) and positive to zero (red).  The values of supercurrent on two sides are expected to be equal in a conventional  JJ. }
    \label{fig:fig_1}
\end{figure}

Our junctions are based on epitaxial superconducting Al thin films grown in-situ on InAs heterostructures by molecular beam epitaxy on a InP substrate followed by a graded buffer layer \cite{shabani_two-dimensional_2016, wickramasinghe_transport_2018, strickland_controlling_2022}. Typically, the critical field of thin film of Al is greater than 1~T. \cref{fig:fig_1}(a) shows a general schematic of our planar JJs. We study junctions with varying  superconducting widths from $W_\text{S}=\SI{0.15}{\micro m}$ to $W_\text{S}=\SI{1}{\micro m}$. All the junctions are $W=\SI{4}{\micro m}$ wide and are fabricated using a transene selective wet etching of Al. \cref{fig:fig_1}(b) shows a false colored scanning electron microscopy (SEM) image of  a typical $L=\SI{150}{\nano m}$ long junction with superconducting width of $W_\text{S}=\SI{1}{\micro m}$. The Al induced gap in our junctions is about $\Delta=\SI{220}{\micro eV}$ estimated from critical temperature, $T_\text{c}$. The semiconductor-superconductor transparency of our junctions are reported in our previous works \cite{mayer_superconducting_2019, mayer_gate_2020, dartiailh_missing_2021} and can host modes with near unity transparency.  All the measurements in this study are performed at $T \approx \SI{30}{mK}$ in a  dilution refrigerator equipped with a three-axis vector magnet. As shown in \cref{fig:fig_1}(b), the $z$-axis of the magnet is perpendicular to the sample plane, while $x$ and $y$-axes are in-plane components aligned parallel to the current and junction, respectively.

\cref{fig:fig_1}(c) presents the differential resistance as a function of the bias current and applied out-of-plane magnetic field for the junction JJ1 with $W_\text{S}=$ $\SI{0.6}{\micro m}$ when the in-plane field is set to zero. The observed Fraunhofer pattern shows a hysteresis due to heating effects when bias is swept through zero~\cite{courtois_origin_2008, mayer_superconducting_2019}. The critical current of the hot electrons branch, where the bias goes from high bias to zero, is clearly smaller than the critical current of the cold electrons branch going from zero to high bias. This is due to the difference between the effective electronic temperature of the hot and cold electrons branches before the transition to or out of the superconducting state. Such a hysteretic behavior leads to different values of critical current on each side,  as can be observed in \cref{fig:fig_1}(d) for JJ1, and has to be avoided for accurate SDE  measurements. In addition, \cref{fig:fig_1}(c) shows that the cold electron branch exhibits a broad switching distribution near the Fraunhofer maximum with several premature switching events.  For the rest of this study, we therefore only derive the values of the critical current from the hot electrons branch, going from high bias to zero bias as shown in \cref{fig:fig_1}(e). These two values, i.e.\ the positive and negative retrapping currents, are expected to be equal in magnitude in reciprocal measurements of a conventional device without presence of in-plane magnetic field.

\begin{figure}[ht!]
    \centerline{\includegraphics[scale=0.5]{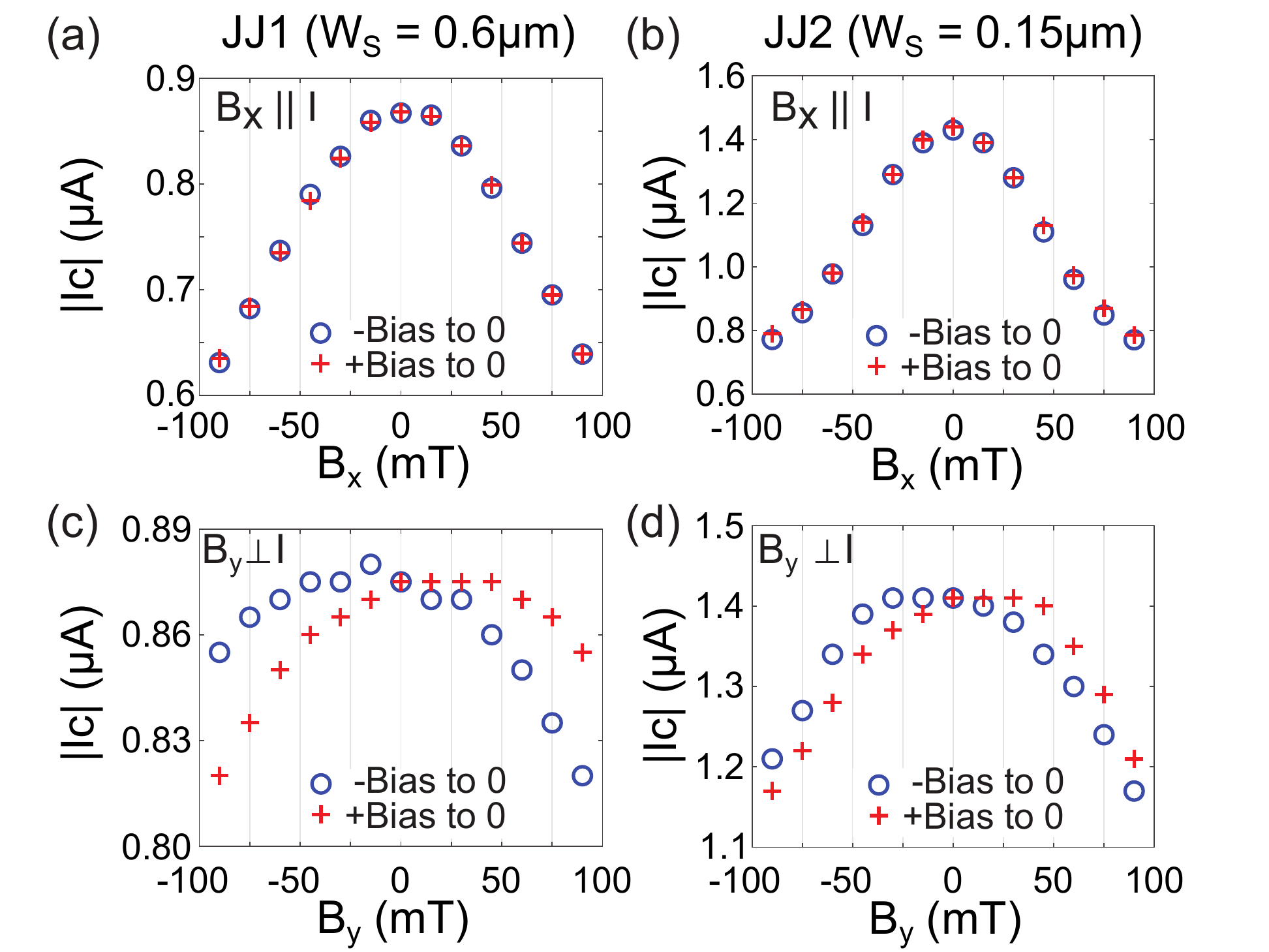}}
    \caption{\textbf{Low in-plane magnetic field dependence.} Absolute value of the critical currents as a function of the in-plane magnetic field for (a),(c) Josephson junction 1 (JJ1) and (b),(d) JJ2, with superconducting contact widths $W_\text{S}$ indicated. (a--b) and (c--d) correspond to an in-plane magnetic field parallel and perpendicular to the current, respectively. The blue circles represent the magnitude of supercurrent when the bias is swept from negative to zero while red cross marks are for bias from positive to zero. The critical current amplitudes when the magnetic field is parallel to the current ($\bm{B}\parallel \hat{\bm{x}}$) are  nearly equal in both directions, indicating a vanishing superconducting diode effect (SDE) [(a) and (b)]. When the magnetic field is perpendicular to the current ($\bm{B}\parallel \hat{\bm{y}}$), the amplitudes of the forward and reverse critical currents are different, signaling the presence of the SDE [(c) and (d)]. Note that nonreciprocity can be observed in both devices. }
    \label{fig:fig_2}
\end{figure}

\subsection{Low in-plane field dependence}

By carefully aligning the magnet directions to Josephson junction and eliminating unwanted out-of-plane component of magnetic field ($B_z$), we measure the critical current in the presence of an in-plane magnetic field. \cref{fig:fig_2}(a) and (b) show the measured magnitude of the critical current $|I_\text{c}|$ for JJ1 with $W_\text{S}=$ $\SI{0.6}{\micro m}$ and JJ2 with $W_\text{S}=$ $\SI{0.15}{\micro m}$ when $B_z=$~0 T and the in-plane magnetic field with strength $B_x$ is parallel to the current. Blue circles and red cross marks correspond to measurement of the magnitude of the critical current when the  bias is swept from negative high bias to zero and from positive high bias to zero, respectively. We find that the magnitude of the critical current in both directions is the same and there is no sign of nonreciprocity when the applied in-plane magnetic field is parallel to the current. The absence of SDE when the field is parallel to the current indicates that the dominant SOC in the junctions is of Rashba type, which is in agreement with our previous works~\cite{wickramasinghe_transport_2018, farzaneh_magneto_anisotropic_2022}.

\begin{figure}[htp]
    \centerline{\includegraphics[width=\linewidth]{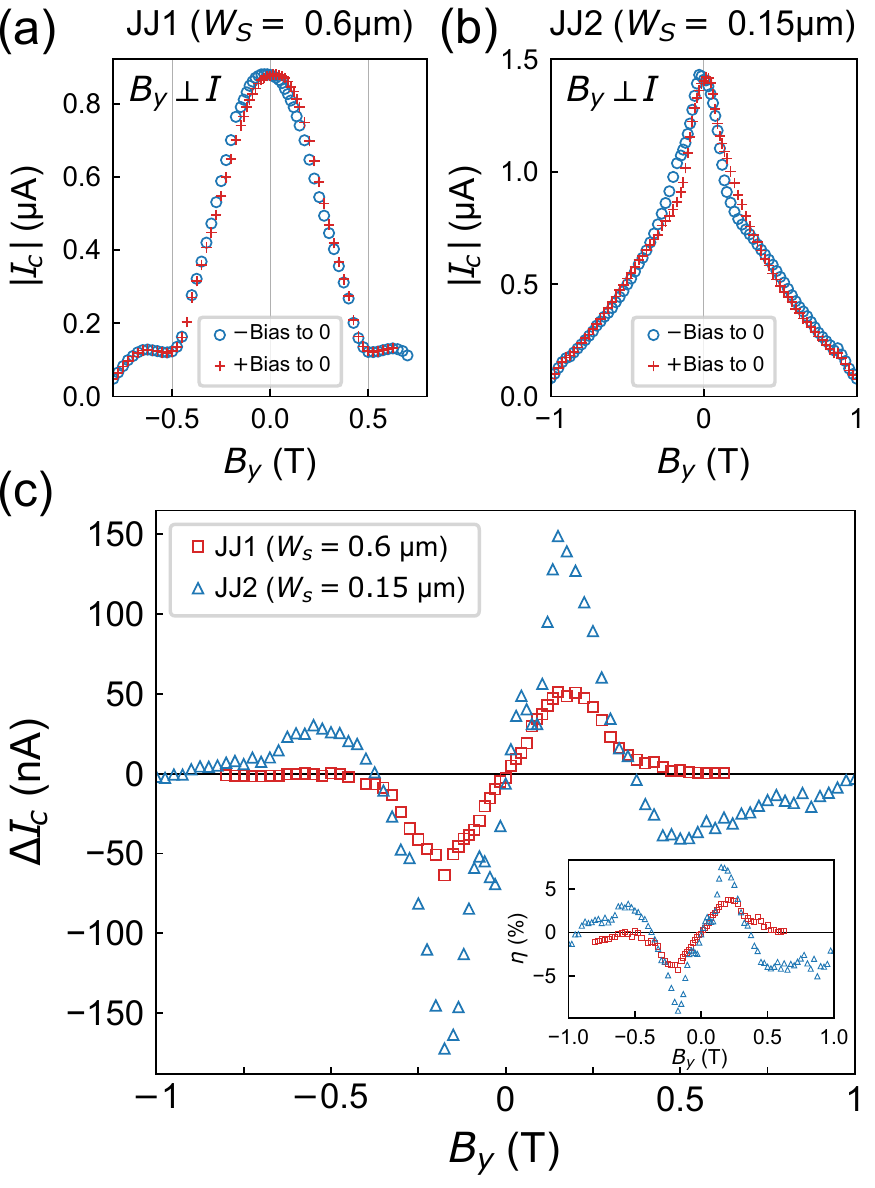}}
    \caption{\textbf{High in-plane magnetic field dependence and sign change.} Absolute value of supercurrent as a function of in-plane magnetic field perpendicular to the current in (a) Josephson junction 1 (JJ1), and (b) JJ2 at high  magnetic fields, with superconducting contact widths $W_\text{S}$ indicated. The blue circles represent the magnitude of supercurrent when the bias is swept from negative to zero while red cross marks are for bias from positive to zero. (c) Difference $\Delta I_\text{c} = I_\text{c}^+ - |I_\text{c}^-| $ between the absolute value of the critical currents measured under positive and negative biases as a function of $B_y$. Red squares and blue triangles correspond to JJ1,  and JJ2, respectively.  Inset: diode efficiency $\eta = (I_\text{c}^+ - |I_\text{c}^-|) / (I_\text{c}^+ + |I_\text{c}^-|)$.}
    \label{fig:fig_q}
\end{figure}

When the in-plane magnetic field is applied perpendicular to the current, in the $y$-direction, we find a difference between the forward and reverse critical currents. \cref{fig:fig_2}(c) and (d) show the dependence of the absolute value of critical current $|I_\text{c}|$ on $B_y$ for JJ1 and JJ2. We observe a clear nonreciprocal behavior, where the critical current is larger for positive than for negative bias when $B_{y}>0$. This behaviour is reversed when the in-plane field direction is flipped to  $B_{y}<0$, in agreement with the theoretically expected symmetry relation $I_\text{c}^+(B_y)=|I_\text{c}^-(-B_y)|$. Details of the experimental measurements and analyses are given in Supplementary Note 1. We extract $I_\text{c}$ at each in-plane magnetic field from the maximum of the Fraunhofer pattern at that field (Fig. S3 and S4 for more details). The same measurements were done on three additional devices with $W_{S}=0.4$, $0.8$ and $\SI{1}{\micro m}$ and showed the same results, as presented in Fig. S5 and S6 of Supplementary Note 1. 
Those devices exhibit the same behavior as JJ1 and JJ2. In all cases we observe a shift, $B_*$ in the magnetic fields at which the critical currents reach their maximum values when $B_y\perp I$. The shift is positive for the critical current corresponding to positive bias and negative for the case of negative bias. In JJ1 and JJ2, the magnitude of shift on  both  negative and positive sides are the same and equal to $|B_*|$=15 mT. Although the value of $|I_\text{c}|$ at few magnetic fields are very close to each other, but from the zoom-in plots of the data presented in Fig. S7, S8 and S9 for JJ1, we can see that the maximum of $|I_\text{c}|$ is at 15 mT for both positive and negative fields (dashed  black line in both figures).

This observed shift is captured by our numerical tight-binding simulations (see details in Supplementary Note 2). The result of the tight-binding simulation in Fig. S13(a) for a junction with $W_\text{S}=\SI{0.15}{\micro m}$ clearly shows the superconducting diode effect in the splitting of $I_\text{c}^\pm$, as well as the symmetry with respect to the sign of $B_y$. Moreover, $B_*$ indicates the presence of SOC in the junction and obeys the symmetry relation, $B_*(\alpha)=-B_*(-\alpha)$, where $\alpha$ denotes the strength of the Rashba SOC. (We neglect Dresselhaus SOC for simplicity, as Rashba SOC is typically dominant in this system \citep{farzaneh_magneto_anisotropic_2022}.) As an illustration, we perform numerical calculations of the magnetic field dependence of the critical currents for JJ2 and different values of the Rashba SOC strength. By tracking the fields at which the numerically calculated critical current maxima occur, we extract the $\alpha$-dependence of $B_*$ [see Fig. S13(b)]. The black, dashed line is just a linear fit to guide the eye. Comparing the field-shift value extracted from the experimental and the corresponding numerical simulations, we estimate the Rashba SOC strength in device JJ2 to be about 10~meV~nm. This value is in overall agreement with values of $\alpha$ in InAs extracted through weak antilocalization measurements~\cite{wickramasinghe_transport_2018, farzaneh_magneto_anisotropic_2022}. Although Fig. S13(b) was specifically computed for JJ2, from the fabrication process and similar composition, we expect all the samples to have similar SOC strengths.

Complementary to the numerical simulations, we provide approximate analytical expressions for the normalized critical currents at low field (see Supplementary Note 2 for detailed information),
\begin{equation}\label{ic-pm}
    \frac{|I_\text{c}^{\pm}|}{I_0}=1-b\left[1\pm c\; {\rm sgn}(B_y \mp B_{\ast})\right](B_y \mp B_{\ast})^2,
\end{equation}
where $I_0$ is the maximum absolute value of the critical current, $b=(g^\ast \mu_\text{B}/4E_\text{T})^2$, $c=k_\text{so}/k_\text{F}$, and $B_{\ast}\approx (1-\tau)^{1/4}(c/\sqrt{b})$ (with $\tau$ as the junction transparency) is the magnitude of the field at which $I_\text{c}$ is maximum. Here $g^\ast$ is the effective g-factor, $\mu_\text{B}$ the Bohr magneton, $k_\text{F}$ the Fermi wavevector, $E_\text{T}=\hbar v_\text{F}/(2 L)$ the Thouless energy, $v_\text{F}$ the Fermi velocity, and $k_\text{so}=\alpha m^\ast/\hbar^2$, with $m^\ast$ representing the effective mass. As discussed in Supplementary Note 2, equation (\ref{ic-pm})  was obtained in the limits $L\ll \xi_0$ where $\xi_0$ is the superconducting coherence length and $W_\text{S}\rightarrow \infty$, assuming the Zeeman interaction is sizable in the N region only, and again neglecting Dresselhaus SOC. Therefore it is not in quantitative agreement with finite $W_\text{S}$ in experimental devices. However, Eq.~(\ref{ic-pm}) can provide a qualitative description of the main trends exhibited by the critical currents. In fact, Eq.~(\ref{ic-pm}) reproduces well the functional behavior of the experimental data at low field. Fig.\ S10 shows the experimental data of all the junctions studied fitted to Eq.~(\ref{ic-pm}) using $b$, $c$ and $B_{\ast}$ as fitting parameters, while Fig.\ S11 shows that $W_\text{S}$ is not predictive of the $B_*$ observed in the presented devices. According to the simplified analytical model, the asymptotic behavior of $B_*$ at low magnetic fields does not only depend on the SOC strength but also on other system parameters, like the junction transparency (see Supplementary Note 2 and Ref.~\onlinecite{Costa2023:arXiv}). Hence devices exhibiting larger values of $B_*$ (see bottom row of Fig. S10) may still have similar SOC strength with lower transparency.

Our experimental data together with numerical simulation suggest that the observed SDE originates from the finite-momentum Cooper pairing induced by the shift of the Fermi contours when the Zeeman interaction and the Rashba SOC coexist as illustrated in Fig. S13(c)-(e). This picture implicitly follows from the microscopic model used in the numerical simulations, which in turn are able to explain the trends observed in the experimental data. Note that the observed SDE depends on both the magnetic field strength and direction. Therefore, the non-intrisic contributions to the SDE originating from gate-dependent effective disorder in the superconducting electrodes \cite{Lo2013:SR}, which are independent of the magnetic field, can be ruled out in our devices. Vortex asymmetric motion, another mechanism that may induce a non-reciprocal behavior \cite{Lustikova2018:NC,Zhang2020:NC,Masuko2022:npjQM}, can also be disregarded as the origin of the SDE in our samples. Indeed, vortex asymmetric motion is expected to be relevant near the superconducting transition when the temperature and/or the applied field are close to their superconducting critical values. However, the SDE here reported is finite at magnetic fields as low as few mT
(i.e., at fields much smaller than the critical field $B_\text{c} \approx 1.6$~T) and temperature of 30~mK (well
below $T_\text{c} \approx 1.5$~K).

In the regime $E_\text{Z}\ll \alpha k_\text{F}$, where $E_\text{Z}$, $\alpha$, and $k_\text{F}$ denote the Zeeman energy, the Rashba SOC strength and the Fermi wave vector, respectively, the Fermi contours in the N region can be approximated as,

\begin{equation}
    k_\lambda = -\lambda k_\text{so}+\sqrt{k_\text{F}^2+k_\text{so}^2+\lambda \kappa^2 \sin(\varphi-\theta)},
\end{equation}
where $\lambda=\pm 1$, $\kappa=\sqrt{2m^\ast E_\text{Z}/\hbar^2}$, and $\theta$ and $\varphi$ determine the directions of the wave vector and magnetic field with respect to the $x$-axis, respectively. The $x$-component of the total momentum of the pairs is,
\begin{equation}\label{eq:q}
    q\approx \frac{\kappa^2\sin\varphi}{\sqrt{k_\text{F}^2+k_\text{so}^2}}
\end{equation}
and the Cooper pair wave function across the junction can be approximated as,
\begin{equation}
 |\psi\rangle=|\uparrow\downarrow\rangle\; \text{e}^{\text{i} q x}+|\downarrow\uparrow\rangle\; \text{e}^{-\text{i} q x}
\end{equation}
and can be rewritten in terms of singlet, $|S\rangle = |\uparrow\downarrow\rangle+|\downarrow\uparrow\rangle$ and triplet, $|T\rangle=|\uparrow\downarrow\rangle-|\downarrow\uparrow\rangle$ components \cite{Eschrig2015:RPP},
\begin{equation}\label{eq:psi}
    |\psi\rangle=\cos(q x)|S\rangle+\text{i}\sin(q x)|T\rangle.
\end{equation}
For $E_\text{Z}\ll \alpha k_\text{F}$, an inversion of the magnetic field orientation reverses the direction of the Fermi contours shift without affecting the spin orientation. Therefore, the coexistence of the singlet and triplet components in the presence of SOC breaks the inversion symmetry of the wave function with respect to the magnetic field direction, resulting in a non-reciprocal response with distinct forward and reverse critical currents. However, the SDE vanishes when the magnetic field is oriented along the $x$-axis (see \cref{fig:fig_2}(a--b)) for in this case $\varphi =0$ and $q=0$ in Eq.~(\ref{eq:q}).

\subsection{High in-plane field dependence}

We further investigate the nonreciprocity of the critical currents at higher in-plane magnetic fields perpendicular to the current ($B_{y}$) in the devices JJ1 and JJ2. \cref{fig:fig_q}(a) and (b) show the absolute value of the critical currents for each junction as a function of $B_{y}$. A dip and peak in $|I_\text{c}|$ of JJ1 is observed around $B_{y} \sim \SI{0.6}{T}$. Previous studies have suggested such a behavior can be related to the closing and reopening  of the superconducting gap \cite{pientka_topological_2017,dartiailh_phase_2021} and a topological phase transition. Our numerical simulations exhibit a phase transition at magnetic field near 0.6~T for $W_\text{S} = \SI{0.6}{\micro m}$ as shown in Fig. S14. In contrast, JJ2 data does not show any peak or dip in the supercurrent in \cref{fig:fig_q}(b). Numerical simulations also do not show a phase transition for the ground state of JJ2 with $W_\text{S} = \SI{0.15}{\micro m}$ below 1~T as shown in Fig. S15.

\begin{figure}[ht!]
    \centerline{\includegraphics[scale=0.48]{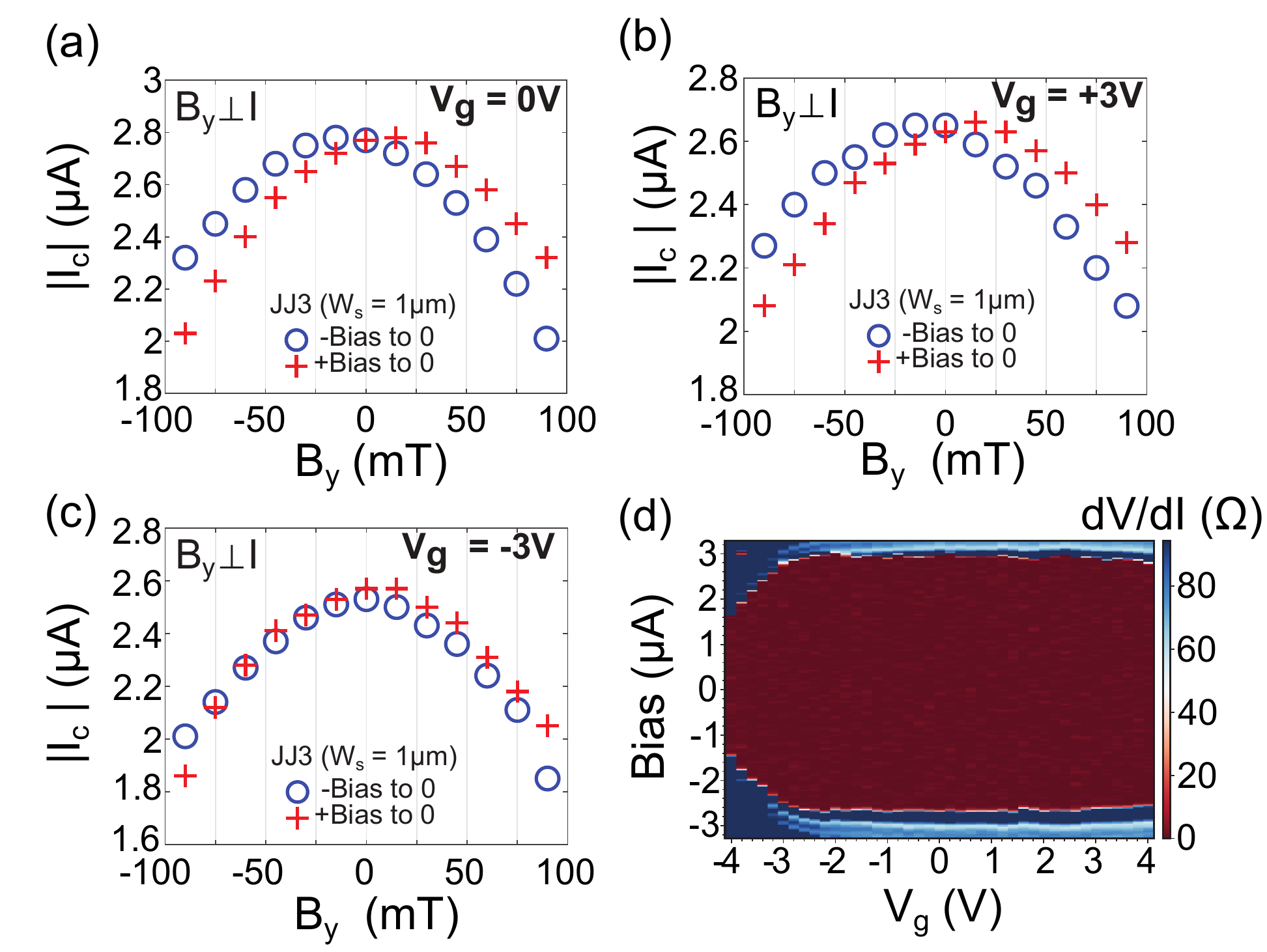}}
    \caption{\textbf{Gate dependence at low in-plane magnetic field.} Absolute value of the critical currents $I_\text{c}$ as a function of the in-plane magnetic field perpendicular to the current $\bm{B}\parallel \hat{\bm{y}}$) for the gated Josephson junction (JJ3) with superconducting contact width $W_\text{S}=$ $\SI{1.0}{\micro m}$ at three different gate voltages: (a) $V_g = \SI{0}{V}$, (b) $V_g = \SI{+3}{V}$, and (c) $V_g = \SI{-3}{V}$. The blue circles represent the magnitude of supercurrent when the bias is swept from negative to zero while red cross marks are for bias from positive to zero. The nonreciprocity can be observed the when the gate voltage is $V_g = 0 , +3V$, indicating the presence of the superconducting diode effect (SDE) [(a) and (b)]. When the gate voltage is $V_g = -3V$, the amplitudes of the forward and reverse critical currents are almost the same, indicating a vanishing  of the SDE [(c)]. (d) Differential resistance of JJ3 as a function of the bias current and applied gate voltage at zero field. Here, the bias was swept from negative to positive values: the difference in positive and negative critical current observed is a hysteretic artifact, see \cref{fig:fig_1}(c--d) and the surrounding discussion.}
    \label{fig:fig_gate}
\end{figure}

\cref{fig:fig_q}(c) plots the difference between the absolute values of the critical currents for positive and negative biases  $\Delta I_\text{c} = I_\text{c}^+ - |I_\text{c}^-|$,  as a function of $B_{y}$.  The results evidence the anti-symmetric character of $\Delta I_\text{c}$, which for both junctions changes its sign when the magnetic field $B_{y}$ is inverted. However, $\Delta I_\text{c}$ also exhibit zeros at certain values of $B_y$, across which sign reversals not related to magnetic field inversion are observed. This is particularly apparent for the device JJ2 (blue symbols) at fields $B_y\approx \pm 0.35$~T, as shown in \cref{fig:fig_q}(c).

From a comparison between the experimental results and the numerical simulations, we identify two possible mechanisms responsible for the zeros of $\Delta I_\text{c}$ and their associated SDE sign reversals. According to Eq.~(\ref{ic-pm}), the SOC induces an asymmetry in the critical currents under the magnetic field inversion with respect to $B_*$, with $|I_\text{c}^\pm(B_*+\delta B)|\neq |I_\text{c}^\pm(B_*-\delta B)|$. This asymmetry is apparent in \cref{fig:fig_2}(c)-(d) and \cref{fig:fig_q}(a)-(b). The coexistence of a finite magnetic shift, $B_*$, and a strong SOC-induced critical current asymmetry can cause $|I_\text{c}^+|$ and $|I_\text{c}^-|$ to cross at a finite magnetic field and produce a sign reversal in the SDE without involving 0--$\pi$-like transitions. This situation is apparent in JJ2 from \cref{fig:fig_q}(b) and (c), where a critical current crossing and corresponding sign reversal of $\Delta I_\text{c}$ at $B_y\approx 0.35$~T are observed, respectively. The numerical simulations are in good agreement with the experimental data of JJ2, predicting a critical current crossing at $B_y=0.4$~T, which is unrelated to the 0--$\pi$ transition at $B_y\approx 1$~T (see Fig. S15 in Supplementary Note 2).

As discussed above, the SDE originates from finite-momentum Cooper pairing qualitatively described by a wave function lacking inversion symmetry with respect to $B_y$ when both Rashba SOC and Zeeman interaction are present. However, it follows from Eq.~(\ref{eq:psi}) that the inversion symmetry with respect to $B_y$ is reestablished when either the singlet or triplet component vanishes at the S/N interfaces located at $x=0$ and $x=L$, i.e., when $|q| L = n\pi/2$, where $n$ is an integer number. Therefore, junctions with $L\ll\xi_0\ll W_\text{S}$ exhibit zeros of $\Delta I_\text{c}$ when,
\begin{equation}\label{By-condition}
    B_y\approx n \frac{\pi}{g^\ast \mu_\text{B}}E_\text{T}\left(1+\frac{k_\text{so}^2}{2k_\text{F}^2}\right)\,r_l.
\end{equation}
The re-scaling factor $r_l=L/(2 W_\text{S} + L)$ has been introduced to account for the fact that the Zeeman field is likely present over the whole system and not only in the semiconductor region.

The zeros (and their associated SDE sign reversals) corresponding to odd integers in Eq.~(\ref{By-condition}), say $n=(2m+1)$ (with $m$ an integer), can be associated with 0--$\pi$-like transitions the junction would experience close to equilibrium. Indeed, in the absence of currents, the superconducting phase difference self-tunes to a value $\phi_\text{GS}$ (referred to as the ground-state phase difference) that minimizes the free energy of the system. For $\cos(q L)>0$ the singlet component of the wave function at the two superconducting leads has the same sign, indicating that $\phi_\text{GS}=0$. However, when $\cos(q W_L)=0$ [i.e., $q L = (2m+1)\pi/2$], the ground-state phase jumps from 0 to $\pi$ and the singlet at the two superconducting leads acquire opposite signs for $\cos(q L)>0$. Therefore, SDE sign reversals corresponding to odd values of $n$ in Eq.~(\ref{By-condition}) are associated to 0--$\pi$-like (or $\pi$--0-like) transitions, while additional sign reversals are expected to occur between 0--$\pi$ and $\pi$--0-like transitions, when $n$ is even. The 0--$\pi$-like ground-state phase jump has been identified as a possible signature of topological phase transitions in planar JJs \cite{pientka_topological_2017,dartiailh_phase_2021}. Hence, the nodes of $\Delta I_\text{c}$ corresponding to odd $n$  may indirectly signal a transition into the topological superconducting state. However, such a signature is not conclusive, especially in JJs with narrow superconducting leads, where ground-state phase jumps are not necessarily associated to topological phase transitions \cite{Setiawan2019:PRB,Pakizer2021:PRB}.

The numerical simulations for devices JJ1 and JJ2 reveal 0--$\pi$-like jumps of the ground-state phase at $B_y\approx\pm 0.6$~T  and $B_y\approx\pm 1$~T, respectively [see Figs. S13 and S14(b) in Supplementary Note 2], suggesting that if the $\Delta I_\text{c}$ changes sign at higher fields [see insets in \cref{fig:fig_q}(c)] they could be associated to 0--$\pi$-like transitions with $n=1$. However, we find that the measured current difference, $\Delta I_\text{c}$, is too small in experiment and is difficult to conclusively establish the existence of these sign reversals in range of 0.6~T to 1~T.

\subsection{Gate dependence}

To further investigate the effect of SOC on the SDE, we make a gated device JJ3 with $W_\text{S}=$ $\SI{1.0}{\micro m}$ by depositing 60 nm Al$_2$O$_3$ followed by 5/40nm Cr/Au on the junction. The thickness of Al in JJ3 is 8nm and the QW has the exact same structure as the QW in other junctions in this study. \cref{fig:fig_gate}(a--c) shows the measured magnitude of the critical current of JJ3 at three different gate voltages when $B_z = 0$ and the in-plane magnetic field is perpendicular to the current ($B_y$). A clear nonreciprocal behavior can be seen when $V_g$ = 0 and +3V. However, the SDE is negligible when $V_g = \SI{-3}{V}$. It has been reported that in InSb nanowires, the SDE drastically depends on the applied gate and orientation of magnetic field and can be suppressed in the absence of SOC in the system  \cite{mazur_gate-tunable_2022}. In previous studies of SOC effects in our system with gated Hall bar measurements, we reported that the Rashba strength can be tuned by gate voltage and is smaller at low densities\cite{farzaneh_magneto_anisotropic_2022, wickramasinghe_transport_2018}. \cref{fig:fig_gate}(d) shows the differential resistance of JJ3 as a function of bias and gate voltage at zero magnetic field. At $V_g= \SI{-3}{V}$ the density is much lower than $V_g= $0 and +3V. The absence of SDE in \cref{fig:fig_gate}(c) suggests that  at $V_g= -3$V the Rashba parameter is significantly smaller than the SOC strength at zero and positive gate voltages.

\section{Conclusion}

In summary, we have studied the superconducting diode effect in epitaxial InAs/Al Josephson junctions with different superconducting width and showed that the SDE depends on the orientation of the applied in-plane magnetic field in the system. By measuring the supercurrent of the junction, we observe SDE only when the in-plane field is perpendicular to the current. We observe a shift in magnetic field yielding the maximum critical current and obtain an analytical expression describing the critical current behavior at low magnetic field. We propose a method for estimating the Rashba parameter from the measurement of the magnetic field shift of the SDE and numerical simulations. The results are in good agreement with values previously reported for our system. We also measure the SDE  at high magnetic fields and observe a sign change in the $\Delta I_\text{c}$ of the $W_\text{S}=$ $\SI{0.15}{\micro m}$ junction at $B_y\approx\pm 0.35$~T. Using our Tight binding  simulation, we conclude that this sign change is not necessarily an indicator of 0--$\pi$ or topological transitions in the system. By measuring the SDE in a gated junction at three different gate voltages, we showed that the SDE   strongly depends on the applied gate voltage and the SOC strength in our system.

\section{Methods}
Wafers are grown by molecular beam epitaxy.  Devices are fabricated using a combination of wet etching and deposition techniques after patterning polymethyl methacrylate (PMMA) via electron beam lithography.  Device mesa features are defined by a deep wet etch with 85\% concentrated phosphoric acid, 30\% concentrated hydrogen peroxide, and deionized water in a volumetric ratio of 1:1:40 after selectively etching the aluminum top layer with Transene Aluminum Etchant Type D.  Junction weak links and smaller device features are defined by a subsequent aluminum etch.  Gated devices subsequently undergo dielectric deposition of aluminum oxide via atomic layer deposition, and titanium/gold gates are deposited via electron beam evaporation.  D.c.\ measurements are performed in a dilution refrigerator at a temperature of around \SI{30}{mK} using standard low-frequency lock-in amplification techniques with excitation currents of at most \SI{10}{nA} and frequencies of around \SI{17}{Hz}.  Magnetic field is generated by a three-axis superconducting vector magnet.

\section{Data Availability}
The transport data generated in relation to this study are available in Zenodo (\href{https://doi.org/10.5281/zenodo.10810819}{doi:10.5281/zenodo.10810819}) \cite{datarepo}.

\section{Code Availability}
All code related to this work is available from the corresponding author upon request.

\bibliography{Ref}

\providecommand{\noopsort}[1]{}\providecommand{\singleletter}[1]{#1}%
\begin{thebibliography}{42}%
\makeatletter
\providecommand \@ifxundefined [1]{%
 \@ifx{#1\undefined}
}%
\providecommand \@ifnum [1]{%
 \ifnum #1\expandafter \@firstoftwo
 \else \expandafter \@secondoftwo
 \fi
}%
\providecommand \@ifx [1]{%
 \ifx #1\expandafter \@firstoftwo
 \else \expandafter \@secondoftwo
 \fi
}%
\providecommand \natexlab [1]{#1}%
\providecommand \enquote  [1]{``#1''}%
\providecommand \bibnamefont  [1]{#1}%
\providecommand \bibfnamefont [1]{#1}%
\providecommand \citenamefont [1]{#1}%
\providecommand \href@noop [0]{\@secondoftwo}%
\providecommand \href [0]{\begingroup \@sanitize@url \@href}%
\providecommand \@href[1]{\@@startlink{#1}\@@href}%
\providecommand \@@href[1]{\endgroup#1\@@endlink}%
\providecommand \@sanitize@url [0]{\catcode `\\12\catcode `\$12\catcode
  `\&12\catcode `\#12\catcode `\^12\catcode `\_12\catcode `\%12\relax}%
\providecommand \@@startlink[1]{}%
\providecommand \@@endlink[0]{}%
\providecommand \url  [0]{\begingroup\@sanitize@url \@url }%
\providecommand \@url [1]{\endgroup\@href {#1}{\urlprefix }}%
\providecommand \urlprefix  [0]{URL }%
\providecommand \Eprint [0]{\href }%
\providecommand \doibase [0]{https://doi.org/}%
\providecommand \selectlanguage [0]{\@gobble}%
\providecommand \bibinfo  [0]{\@secondoftwo}%
\providecommand \bibfield  [0]{\@secondoftwo}%
\providecommand \translation [1]{[#1]}%
\providecommand \BibitemOpen [0]{}%
\providecommand \bibitemStop [0]{}%
\providecommand \bibitemNoStop [0]{.\EOS\space}%
\providecommand \EOS [0]{\spacefactor3000\relax}%
\providecommand \BibitemShut  [1]{\csname bibitem#1\endcsname}%
\let\auto@bib@innerbib\@empty
\bibitem [{\citenamefont {Ando}\ \emph {et~al.}(2020)\citenamefont {Ando},
  \citenamefont {Miyasaka}, \citenamefont {Li}, \citenamefont {Ishizuka},
  \citenamefont {Arakawa}, \citenamefont {Shiota}, \citenamefont {Moriyama},
  \citenamefont {Yanase},\ and\ \citenamefont {Ono}}]{ando_observation_2020}%
  \BibitemOpen
  \bibfield  {author} {\bibinfo {author} {\bibfnamefont {F.}~\bibnamefont
  {Ando}}, \bibinfo {author} {\bibfnamefont {Y.}~\bibnamefont {Miyasaka}},
  \bibinfo {author} {\bibfnamefont {T.}~\bibnamefont {Li}}, \bibinfo {author}
  {\bibfnamefont {J.}~\bibnamefont {Ishizuka}}, \bibinfo {author}
  {\bibfnamefont {T.}~\bibnamefont {Arakawa}}, \bibinfo {author} {\bibfnamefont
  {Y.}~\bibnamefont {Shiota}}, \bibinfo {author} {\bibfnamefont
  {T.}~\bibnamefont {Moriyama}}, \bibinfo {author} {\bibfnamefont
  {Y.}~\bibnamefont {Yanase}},\ and\ \bibinfo {author} {\bibfnamefont
  {T.}~\bibnamefont {Ono}},\ }\bibfield  {title} {\bibinfo {title} {Observation
  of superconducting diode effect},\ }\href
  {https://doi.org/10.1038/s41586-020-2590-4} {\bibfield  {journal} {\bibinfo
  {journal} {Nature}\ }\textbf {\bibinfo {volume} {584}},\ \bibinfo {pages}
  {373} (\bibinfo {year} {2020})}\BibitemShut {NoStop}%
\bibitem [{\citenamefont {Baumgartner}\ \emph
  {et~al.}(2022{\natexlab{a}})\citenamefont {Baumgartner}, \citenamefont
  {Fuchs}, \citenamefont {Costa}, \citenamefont {Picó-Cortés}, \citenamefont
  {Reinhardt}, \citenamefont {Gronin}, \citenamefont {Gardner}, \citenamefont
  {Lindemann}, \citenamefont {Manfra}, \citenamefont {Faria~Junior},
  \citenamefont {Kochan}, \citenamefont {Fabian}, \citenamefont {Paradiso},\
  and\ \citenamefont {Strunk}}]{baumgartner_effect_2022}%
  \BibitemOpen
  \bibfield  {author} {\bibinfo {author} {\bibfnamefont {C.}~\bibnamefont
  {Baumgartner}}, \bibinfo {author} {\bibfnamefont {L.}~\bibnamefont {Fuchs}},
  \bibinfo {author} {\bibfnamefont {A.}~\bibnamefont {Costa}}, \bibinfo
  {author} {\bibfnamefont {J.}~\bibnamefont {Picó-Cortés}}, \bibinfo {author}
  {\bibfnamefont {S.}~\bibnamefont {Reinhardt}}, \bibinfo {author}
  {\bibfnamefont {S.}~\bibnamefont {Gronin}}, \bibinfo {author} {\bibfnamefont
  {G.~C.}\ \bibnamefont {Gardner}}, \bibinfo {author} {\bibfnamefont
  {T.}~\bibnamefont {Lindemann}}, \bibinfo {author} {\bibfnamefont {M.~J.}\
  \bibnamefont {Manfra}}, \bibinfo {author} {\bibfnamefont {P.~E.}\
  \bibnamefont {Faria~Junior}}, \bibinfo {author} {\bibfnamefont
  {D.}~\bibnamefont {Kochan}}, \bibinfo {author} {\bibfnamefont
  {J.}~\bibnamefont {Fabian}}, \bibinfo {author} {\bibfnamefont
  {N.}~\bibnamefont {Paradiso}},\ and\ \bibinfo {author} {\bibfnamefont
  {C.}~\bibnamefont {Strunk}},\ }\bibfield  {title} {\bibinfo {title} {Effect
  of {Rashba} and {Dresselhaus} spin–orbit coupling on supercurrent
  rectification and magnetochiral anisotropy of ballistic {Josephson}
  junctions},\ }\href {https://doi.org/10.1088/1361-648X/ac4d5e} {\bibfield
  {journal} {\bibinfo  {journal} {Journal of Physics: Condensed Matter}\
  }\textbf {\bibinfo {volume} {34}},\ \bibinfo {pages} {154005} (\bibinfo
  {year} {2022}{\natexlab{a}})}\BibitemShut {NoStop}%
\bibitem [{\citenamefont {Baumgartner}\ \emph
  {et~al.}(2022{\natexlab{b}})\citenamefont {Baumgartner}, \citenamefont
  {Fuchs}, \citenamefont {Costa}, \citenamefont {Reinhardt}, \citenamefont
  {Gronin}, \citenamefont {Gardner}, \citenamefont {Lindemann}, \citenamefont
  {Manfra}, \citenamefont {Faria~Junior}, \citenamefont {Kochan}, \citenamefont
  {Fabian}, \citenamefont {Paradiso},\ and\ \citenamefont
  {Strunk}}]{baumgartner_supercurrent_2022}%
  \BibitemOpen
  \bibfield  {author} {\bibinfo {author} {\bibfnamefont {C.}~\bibnamefont
  {Baumgartner}}, \bibinfo {author} {\bibfnamefont {L.}~\bibnamefont {Fuchs}},
  \bibinfo {author} {\bibfnamefont {A.}~\bibnamefont {Costa}}, \bibinfo
  {author} {\bibfnamefont {S.}~\bibnamefont {Reinhardt}}, \bibinfo {author}
  {\bibfnamefont {S.}~\bibnamefont {Gronin}}, \bibinfo {author} {\bibfnamefont
  {G.~C.}\ \bibnamefont {Gardner}}, \bibinfo {author} {\bibfnamefont
  {T.}~\bibnamefont {Lindemann}}, \bibinfo {author} {\bibfnamefont {M.~J.}\
  \bibnamefont {Manfra}}, \bibinfo {author} {\bibfnamefont {P.~E.}\
  \bibnamefont {Faria~Junior}}, \bibinfo {author} {\bibfnamefont
  {D.}~\bibnamefont {Kochan}}, \bibinfo {author} {\bibfnamefont
  {J.}~\bibnamefont {Fabian}}, \bibinfo {author} {\bibfnamefont
  {N.}~\bibnamefont {Paradiso}},\ and\ \bibinfo {author} {\bibfnamefont
  {C.}~\bibnamefont {Strunk}},\ }\bibfield  {title} {\bibinfo {title}
  {Supercurrent rectification and magnetochiral effects in symmetric
  {Josephson} junctions},\ }\href {https://doi.org/10.1038/s41565-021-01009-9}
  {\bibfield  {journal} {\bibinfo  {journal} {Nature Nanotechnology}\ }\textbf
  {\bibinfo {volume} {17}},\ \bibinfo {pages} {39} (\bibinfo {year}
  {2022}{\natexlab{b}})}\BibitemShut {NoStop}%
\bibitem [{\citenamefont {Bauriedl}\ \emph {et~al.}(2022)\citenamefont
  {Bauriedl}, \citenamefont {Bäuml}, \citenamefont {Fuchs}, \citenamefont
  {Baumgartner}, \citenamefont {Paulik}, \citenamefont {Bauer}, \citenamefont
  {Lin}, \citenamefont {Lupton}, \citenamefont {Taniguchi}, \citenamefont
  {Watanabe}, \citenamefont {Strunk},\ and\ \citenamefont
  {Paradiso}}]{bauriedl_supercurrent_2022}%
  \BibitemOpen
  \bibfield  {author} {\bibinfo {author} {\bibfnamefont {L.}~\bibnamefont
  {Bauriedl}}, \bibinfo {author} {\bibfnamefont {C.}~\bibnamefont {Bäuml}},
  \bibinfo {author} {\bibfnamefont {L.}~\bibnamefont {Fuchs}}, \bibinfo
  {author} {\bibfnamefont {C.}~\bibnamefont {Baumgartner}}, \bibinfo {author}
  {\bibfnamefont {N.}~\bibnamefont {Paulik}}, \bibinfo {author} {\bibfnamefont
  {J.~M.}\ \bibnamefont {Bauer}}, \bibinfo {author} {\bibfnamefont {K.-Q.}\
  \bibnamefont {Lin}}, \bibinfo {author} {\bibfnamefont {J.~M.}\ \bibnamefont
  {Lupton}}, \bibinfo {author} {\bibfnamefont {T.}~\bibnamefont {Taniguchi}},
  \bibinfo {author} {\bibfnamefont {K.}~\bibnamefont {Watanabe}}, \bibinfo
  {author} {\bibfnamefont {C.}~\bibnamefont {Strunk}},\ and\ \bibinfo {author}
  {\bibfnamefont {N.}~\bibnamefont {Paradiso}},\ }\bibfield  {title} {\bibinfo
  {title} {Supercurrent diode effect and magnetochiral anisotropy in few-layer
  {NbSe2}},\ }\href {https://doi.org/10.1038/s41467-022-31954-5} {\bibfield
  {journal} {\bibinfo  {journal} {Nature Communications}\ }\textbf {\bibinfo
  {volume} {13}},\ \bibinfo {pages} {4266} (\bibinfo {year}
  {2022})}\BibitemShut {NoStop}%
\bibitem [{\citenamefont {Wu}\ \emph {et~al.}(2022)\citenamefont {Wu},
  \citenamefont {Wang}, \citenamefont {Xu}, \citenamefont {Sivakumar},
  \citenamefont {Pasco}, \citenamefont {Filippozzi}, \citenamefont {Parkin},
  \citenamefont {Zeng}, \citenamefont {McQueen},\ and\ \citenamefont
  {Ali}}]{wu_field-free_2022}%
  \BibitemOpen
  \bibfield  {author} {\bibinfo {author} {\bibfnamefont {H.}~\bibnamefont
  {Wu}}, \bibinfo {author} {\bibfnamefont {Y.}~\bibnamefont {Wang}}, \bibinfo
  {author} {\bibfnamefont {Y.}~\bibnamefont {Xu}}, \bibinfo {author}
  {\bibfnamefont {P.~K.}\ \bibnamefont {Sivakumar}}, \bibinfo {author}
  {\bibfnamefont {C.}~\bibnamefont {Pasco}}, \bibinfo {author} {\bibfnamefont
  {U.}~\bibnamefont {Filippozzi}}, \bibinfo {author} {\bibfnamefont {S.~S.~P.}\
  \bibnamefont {Parkin}}, \bibinfo {author} {\bibfnamefont {Y.-J.}\
  \bibnamefont {Zeng}}, \bibinfo {author} {\bibfnamefont {T.}~\bibnamefont
  {McQueen}},\ and\ \bibinfo {author} {\bibfnamefont {M.~N.}\ \bibnamefont
  {Ali}},\ }\bibfield  {title} {\bibinfo {title} {The field-free {Josephson}
  diode in a van der {Waals} heterostructure},\ }\href
  {https://doi.org/10.1038/s41586-022-04504-8} {\bibfield  {journal} {\bibinfo
  {journal} {Nature}\ }\textbf {\bibinfo {volume} {604}},\ \bibinfo {pages}
  {653} (\bibinfo {year} {2022})}\BibitemShut {NoStop}%
\bibitem [{\citenamefont {Pal}\ \emph {et~al.}(2022)\citenamefont {Pal},
  \citenamefont {Chakraborty}, \citenamefont {Sivakumar}, \citenamefont
  {Davydova}, \citenamefont {Gopi}, \citenamefont {Pandeya}, \citenamefont
  {Krieger}, \citenamefont {Zhang}, \citenamefont {Date}, \citenamefont {Ju},
  \citenamefont {Yuan}, \citenamefont {Schröter}, \citenamefont {Fu},\ and\
  \citenamefont {Parkin}}]{pal_josephson_2022}%
  \BibitemOpen
  \bibfield  {author} {\bibinfo {author} {\bibfnamefont {B.}~\bibnamefont
  {Pal}}, \bibinfo {author} {\bibfnamefont {A.}~\bibnamefont {Chakraborty}},
  \bibinfo {author} {\bibfnamefont {P.~K.}\ \bibnamefont {Sivakumar}}, \bibinfo
  {author} {\bibfnamefont {M.}~\bibnamefont {Davydova}}, \bibinfo {author}
  {\bibfnamefont {A.~K.}\ \bibnamefont {Gopi}}, \bibinfo {author}
  {\bibfnamefont {A.~K.}\ \bibnamefont {Pandeya}}, \bibinfo {author}
  {\bibfnamefont {J.~A.}\ \bibnamefont {Krieger}}, \bibinfo {author}
  {\bibfnamefont {Y.}~\bibnamefont {Zhang}}, \bibinfo {author} {\bibfnamefont
  {M.}~\bibnamefont {Date}}, \bibinfo {author} {\bibfnamefont {S.}~\bibnamefont
  {Ju}}, \bibinfo {author} {\bibfnamefont {N.}~\bibnamefont {Yuan}}, \bibinfo
  {author} {\bibfnamefont {N.~B.~M.}\ \bibnamefont {Schröter}}, \bibinfo
  {author} {\bibfnamefont {L.}~\bibnamefont {Fu}},\ and\ \bibinfo {author}
  {\bibfnamefont {S.~S.~P.}\ \bibnamefont {Parkin}},\ }\bibfield  {title}
  {\bibinfo {title} {Josephson diode effect from {Cooper} pair momentum in a
  topological semimetal},\ }\href {https://doi.org/10.1038/s41567-022-01699-5}
  {\bibfield  {journal} {\bibinfo  {journal} {Nature Physics}\ }\textbf
  {\bibinfo {volume} {18}},\ \bibinfo {pages} {1228} (\bibinfo {year}
  {2022})}\BibitemShut {NoStop}%
\bibitem [{\citenamefont {Yuan}\ and\ \citenamefont
  {Fu}(2022)}]{yuan_supercurrent_2022}%
  \BibitemOpen
  \bibfield  {author} {\bibinfo {author} {\bibfnamefont {N.~F.~Q.}\
  \bibnamefont {Yuan}}\ and\ \bibinfo {author} {\bibfnamefont {L.}~\bibnamefont
  {Fu}},\ }\bibfield  {title} {\bibinfo {title} {Supercurrent diode effect and
  finite-momentum superconductors},\ }\href
  {https://doi.org/10.1073/pnas.2119548119} {\bibfield  {journal} {\bibinfo
  {journal} {Proceedings of the National Academy of Sciences}\ }\textbf
  {\bibinfo {volume} {119}},\ \bibinfo {pages} {e2119548119} (\bibinfo {year}
  {2022})}\BibitemShut {NoStop}%
\bibitem [{\citenamefont {Mazur}\ \emph {et~al.}(2022)\citenamefont {Mazur},
  \citenamefont {van Loo}, \citenamefont {van Driel}, \citenamefont {Wang},
  \citenamefont {Badawy}, \citenamefont {Gazibegovic}, \citenamefont
  {Bakkers},\ and\ \citenamefont {Kouwenhoven}}]{mazur_gate-tunable_2022}%
  \BibitemOpen
  \bibfield  {author} {\bibinfo {author} {\bibfnamefont {G.~P.}\ \bibnamefont
  {Mazur}}, \bibinfo {author} {\bibfnamefont {N.}~\bibnamefont {van Loo}},
  \bibinfo {author} {\bibfnamefont {D.}~\bibnamefont {van Driel}}, \bibinfo
  {author} {\bibfnamefont {J.-Y.}\ \bibnamefont {Wang}}, \bibinfo {author}
  {\bibfnamefont {G.}~\bibnamefont {Badawy}}, \bibinfo {author} {\bibfnamefont
  {S.}~\bibnamefont {Gazibegovic}}, \bibinfo {author} {\bibfnamefont {E.~P.
  A.~M.}\ \bibnamefont {Bakkers}},\ and\ \bibinfo {author} {\bibfnamefont
  {L.~P.}\ \bibnamefont {Kouwenhoven}},\ }\bibfield  {title} {\bibinfo {title}
  {The gate-tunable {Josephson} diode}} (\bibinfo {year} {2022}),\ \bibinfo
  {note} {preprint at \url{https://arxiv.org/abs/2211.14283}}\BibitemShut
  {NoStop}%
\bibitem [{\citenamefont {Sundaresh}\ \emph {et~al.}(2023)\citenamefont
  {Sundaresh}, \citenamefont {V{\"a}yrynen}, \citenamefont {Lyanda-Geller},\
  and\ \citenamefont {Rokhinson}}]{Sundaresh2023:NC}%
  \BibitemOpen
  \bibfield  {author} {\bibinfo {author} {\bibfnamefont {A.}~\bibnamefont
  {Sundaresh}}, \bibinfo {author} {\bibfnamefont {J.~I.}\ \bibnamefont
  {V{\"a}yrynen}}, \bibinfo {author} {\bibfnamefont {Y.}~\bibnamefont
  {Lyanda-Geller}},\ and\ \bibinfo {author} {\bibfnamefont {L.~P.}\
  \bibnamefont {Rokhinson}},\ }\bibfield  {title} {\bibinfo {title}
  {Diamagnetic mechanism of critical current non-reciprocity in multilayered
  superconductors},\ }\href@noop {} {\bibfield  {journal} {\bibinfo  {journal}
  {Nat. Commun.}\ }\textbf {\bibinfo {volume} {14}},\ \bibinfo {pages} {1628}
  (\bibinfo {year} {2023})}\BibitemShut {NoStop}%
\bibitem [{\citenamefont {Gupta}\ \emph {et~al.}(2023)\citenamefont {Gupta},
  \citenamefont {Graziano}, \citenamefont {Pendharkar}, \citenamefont {Dong},
  \citenamefont {Dempsey}, \citenamefont {Palmstrøm},\ and\ \citenamefont
  {Pribiag}}]{gupta_superconducting_2022}%
  \BibitemOpen
  \bibfield  {author} {\bibinfo {author} {\bibfnamefont {M.}~\bibnamefont
  {Gupta}}, \bibinfo {author} {\bibfnamefont {G.~V.}\ \bibnamefont {Graziano}},
  \bibinfo {author} {\bibfnamefont {M.}~\bibnamefont {Pendharkar}}, \bibinfo
  {author} {\bibfnamefont {J.~T.}\ \bibnamefont {Dong}}, \bibinfo {author}
  {\bibfnamefont {C.~P.}\ \bibnamefont {Dempsey}}, \bibinfo {author}
  {\bibfnamefont {C.}~\bibnamefont {Palmstrøm}},\ and\ \bibinfo {author}
  {\bibfnamefont {V.~S.}\ \bibnamefont {Pribiag}},\ }\bibfield  {title}
  {\bibinfo {title} {Gate-tunable superconducting diode effect in a
  three-terminal {Josephson} device},\ }\href
  {https://doi.org/10.1038/s41467-023-38856-0} {\bibfield  {journal} {\bibinfo
  {journal} {Nature Communications}\ }\textbf {\bibinfo {volume} {14}},\
  \bibinfo {pages} {3078} (\bibinfo {year} {2023})}\BibitemShut {NoStop}%
\bibitem [{\citenamefont {Zhang}\ \emph {et~al.}(2022)\citenamefont {Zhang},
  \citenamefont {Li}, \citenamefont {Aguilar}, \citenamefont {Zhang},
  \citenamefont {Pendharkar}, \citenamefont {Dempsey}, \citenamefont {Lee},
  \citenamefont {Harrington}, \citenamefont {Tan}, \citenamefont {Meyer},
  \citenamefont {Houzet}, \citenamefont {Palmstrom},\ and\ \citenamefont
  {Frolov}}]{zhang_evidence_2022}%
  \BibitemOpen
  \bibfield  {author} {\bibinfo {author} {\bibfnamefont {B.}~\bibnamefont
  {Zhang}}, \bibinfo {author} {\bibfnamefont {Z.}~\bibnamefont {Li}}, \bibinfo
  {author} {\bibfnamefont {V.}~\bibnamefont {Aguilar}}, \bibinfo {author}
  {\bibfnamefont {P.}~\bibnamefont {Zhang}}, \bibinfo {author} {\bibfnamefont
  {M.}~\bibnamefont {Pendharkar}}, \bibinfo {author} {\bibfnamefont
  {C.}~\bibnamefont {Dempsey}}, \bibinfo {author} {\bibfnamefont {J.~S.}\
  \bibnamefont {Lee}}, \bibinfo {author} {\bibfnamefont {S.~D.}\ \bibnamefont
  {Harrington}}, \bibinfo {author} {\bibfnamefont {S.}~\bibnamefont {Tan}},
  \bibinfo {author} {\bibfnamefont {J.~S.}\ \bibnamefont {Meyer}}, \bibinfo
  {author} {\bibfnamefont {M.}~\bibnamefont {Houzet}}, \bibinfo {author}
  {\bibfnamefont {C.~J.}\ \bibnamefont {Palmstrom}},\ and\ \bibinfo {author}
  {\bibfnamefont {S.~M.}\ \bibnamefont {Frolov}},\ }\bibfield  {title}
  {\bibinfo {title} {Evidence of $\phi$0-{Josephson} junction from skewed
  diffraction patterns in {Sn}-{InSb} nanowires}} (\bibinfo {year} {2022}),\
  \bibinfo {note} {preprint at
  \url{https://arxiv.org/abs/2212.00199}}\BibitemShut {NoStop}%
\bibitem [{\citenamefont {Margineda}\ \emph {et~al.}(2023)\citenamefont
  {Margineda}, \citenamefont {Crippa}, \citenamefont {Strambini}, \citenamefont
  {Fukaya}, \citenamefont {Mercaldo}, \citenamefont {Cuoco},\ and\
  \citenamefont {Giazotto}}]{margineda2023}%
  \BibitemOpen
  \bibfield  {author} {\bibinfo {author} {\bibfnamefont {D.}~\bibnamefont
  {Margineda}}, \bibinfo {author} {\bibfnamefont {A.}~\bibnamefont {Crippa}},
  \bibinfo {author} {\bibfnamefont {E.}~\bibnamefont {Strambini}}, \bibinfo
  {author} {\bibfnamefont {Y.}~\bibnamefont {Fukaya}}, \bibinfo {author}
  {\bibfnamefont {M.~T.}\ \bibnamefont {Mercaldo}}, \bibinfo {author}
  {\bibfnamefont {M.}~\bibnamefont {Cuoco}},\ and\ \bibinfo {author}
  {\bibfnamefont {F.}~\bibnamefont {Giazotto}},\ }\bibfield  {title} {\bibinfo
  {title} {Sign reversal diode effect in superconducting {Dayem} nanobridges},\
  }\href {https://doi.org/10.1038/s42005-023-01458-9} {\bibfield  {journal}
  {\bibinfo  {journal} {Communications Physics}\ }\textbf {\bibinfo {volume}
  {6}},\ \bibinfo {pages} {1} (\bibinfo {year} {2023})}\BibitemShut {NoStop}%
\bibitem [{\citenamefont {Davydova}\ \emph {et~al.}(2022)\citenamefont
  {Davydova}, \citenamefont {Prembabu},\ and\ \citenamefont
  {Fu}}]{davydova_universal_2022}%
  \BibitemOpen
  \bibfield  {author} {\bibinfo {author} {\bibfnamefont {M.}~\bibnamefont
  {Davydova}}, \bibinfo {author} {\bibfnamefont {S.}~\bibnamefont {Prembabu}},\
  and\ \bibinfo {author} {\bibfnamefont {L.}~\bibnamefont {Fu}},\ }\bibfield
  {title} {\bibinfo {title} {Universal {Josephson} diode effect},\ }\href
  {https://doi.org/10.1126/sciadv.abo0309} {\bibfield  {journal} {\bibinfo
  {journal} {Science Advances}\ }\textbf {\bibinfo {volume} {8}},\ \bibinfo
  {pages} {eabo0309} (\bibinfo {year} {2022})}\BibitemShut {NoStop}%
\bibitem [{\citenamefont {Daido}\ \emph {et~al.}(2022)\citenamefont {Daido},
  \citenamefont {Ikeda},\ and\ \citenamefont {Yanase}}]{daido_intrinsic_2022}%
  \BibitemOpen
  \bibfield  {author} {\bibinfo {author} {\bibfnamefont {A.}~\bibnamefont
  {Daido}}, \bibinfo {author} {\bibfnamefont {Y.}~\bibnamefont {Ikeda}},\ and\
  \bibinfo {author} {\bibfnamefont {Y.}~\bibnamefont {Yanase}},\ }\bibfield
  {title} {\bibinfo {title} {Intrinsic {Superconducting} {Diode} {Effect}},\
  }\href {https://doi.org/10.1103/PhysRevLett.128.037001} {\bibfield  {journal}
  {\bibinfo  {journal} {Physical Review Letters}\ }\textbf {\bibinfo {volume}
  {128}},\ \bibinfo {pages} {037001} (\bibinfo {year} {2022})}\BibitemShut
  {NoStop}%
\bibitem [{\citenamefont {Ilic}\ and\ \citenamefont
  {Bergeret}(2022)}]{ilic_theory_2022}%
  \BibitemOpen
  \bibfield  {author} {\bibinfo {author} {\bibfnamefont {S.}~\bibnamefont
  {Ilic}}\ and\ \bibinfo {author} {\bibfnamefont {F.}~\bibnamefont
  {Bergeret}},\ }\bibfield  {title} {\bibinfo {title} {Theory of the
  {Supercurrent} {Diode} {Effect} in {Rashba} {Superconductors} with
  {Arbitrary} {Disorder}},\ }\href
  {https://doi.org/10.1103/PhysRevLett.128.177001} {\bibfield  {journal}
  {\bibinfo  {journal} {Physical Review Letters}\ }\textbf {\bibinfo {volume}
  {128}},\ \bibinfo {pages} {177001} (\bibinfo {year} {2022})}\BibitemShut
  {NoStop}%
\bibitem [{\citenamefont {He}\ \emph {et~al.}(2022)\citenamefont {He},
  \citenamefont {Tanaka},\ and\ \citenamefont
  {Nagaosa}}]{he_phenomenological_2022}%
  \BibitemOpen
  \bibfield  {author} {\bibinfo {author} {\bibfnamefont {J.~J.}\ \bibnamefont
  {He}}, \bibinfo {author} {\bibfnamefont {Y.}~\bibnamefont {Tanaka}},\ and\
  \bibinfo {author} {\bibfnamefont {N.}~\bibnamefont {Nagaosa}},\ }\bibfield
  {title} {\bibinfo {title} {A phenomenological theory of superconductor
  diodes},\ }\href {https://doi.org/10.1088/1367-2630/ac6766} {\bibfield
  {journal} {\bibinfo  {journal} {New Journal of Physics}\ }\textbf {\bibinfo
  {volume} {24}},\ \bibinfo {pages} {053014} (\bibinfo {year}
  {2022})}\BibitemShut {NoStop}%
\bibitem [{\citenamefont {Costa}\ \emph
  {et~al.}(2023{\natexlab{a}})\citenamefont {Costa}, \citenamefont
  {Baumgartner}, \citenamefont {Reinhardt}, \citenamefont {Berger},
  \citenamefont {Gronin}, \citenamefont {Gardner}, \citenamefont {Lindemann},
  \citenamefont {Manfra}, \citenamefont {Fabian}, \citenamefont {Kochan},
  \citenamefont {Paradiso},\ and\ \citenamefont {Strunk}}]{Costa2022:arxiv}%
  \BibitemOpen
  \bibfield  {author} {\bibinfo {author} {\bibfnamefont {A.}~\bibnamefont
  {Costa}}, \bibinfo {author} {\bibfnamefont {C.}~\bibnamefont {Baumgartner}},
  \bibinfo {author} {\bibfnamefont {S.}~\bibnamefont {Reinhardt}}, \bibinfo
  {author} {\bibfnamefont {J.}~\bibnamefont {Berger}}, \bibinfo {author}
  {\bibfnamefont {S.}~\bibnamefont {Gronin}}, \bibinfo {author} {\bibfnamefont
  {G.~C.}\ \bibnamefont {Gardner}}, \bibinfo {author} {\bibfnamefont
  {T.}~\bibnamefont {Lindemann}}, \bibinfo {author} {\bibfnamefont {M.~J.}\
  \bibnamefont {Manfra}}, \bibinfo {author} {\bibfnamefont {J.}~\bibnamefont
  {Fabian}}, \bibinfo {author} {\bibfnamefont {D.}~\bibnamefont {Kochan}},
  \bibinfo {author} {\bibfnamefont {N.}~\bibnamefont {Paradiso}},\ and\
  \bibinfo {author} {\bibfnamefont {C.}~\bibnamefont {Strunk}},\ }\bibfield
  {title} {\bibinfo {title} {Sign reversal of the {Josephson} inductance
  magnetochiral anisotropy and 0–π-like transitions in supercurrent
  diodes},\ }\href {https://doi.org/10.1038/s41565-023-01451-x} {\bibfield
  {journal} {\bibinfo  {journal} {Nature Nanotechnology}\ }\textbf {\bibinfo
  {volume} {18}},\ \bibinfo {pages} {1266} (\bibinfo {year}
  {2023}{\natexlab{a}})}\BibitemShut {NoStop}%
\bibitem [{\citenamefont {Turini}\ \emph {et~al.}(2022)\citenamefont {Turini},
  \citenamefont {Salimian}, \citenamefont {Carrega}, \citenamefont {Iorio},
  \citenamefont {Strambini}, \citenamefont {Giazotto}, \citenamefont {Zannier},
  \citenamefont {Sorba},\ and\ \citenamefont {Heun}}]{turini_josephson_2022}%
  \BibitemOpen
  \bibfield  {author} {\bibinfo {author} {\bibfnamefont {B.}~\bibnamefont
  {Turini}}, \bibinfo {author} {\bibfnamefont {S.}~\bibnamefont {Salimian}},
  \bibinfo {author} {\bibfnamefont {M.}~\bibnamefont {Carrega}}, \bibinfo
  {author} {\bibfnamefont {A.}~\bibnamefont {Iorio}}, \bibinfo {author}
  {\bibfnamefont {E.}~\bibnamefont {Strambini}}, \bibinfo {author}
  {\bibfnamefont {F.}~\bibnamefont {Giazotto}}, \bibinfo {author}
  {\bibfnamefont {V.}~\bibnamefont {Zannier}}, \bibinfo {author} {\bibfnamefont
  {L.}~\bibnamefont {Sorba}},\ and\ \bibinfo {author} {\bibfnamefont
  {S.}~\bibnamefont {Heun}},\ }\bibfield  {title} {\bibinfo {title} {Josephson
  {Diode} {Effect} in {High}-{Mobility} {InSb} {Nanoflags}},\ }\href
  {https://doi.org/10.1021/acs.nanolett.2c02899} {\bibfield  {journal}
  {\bibinfo  {journal} {Nano Letters}\ }\textbf {\bibinfo {volume} {22}},\
  \bibinfo {pages} {8502} (\bibinfo {year} {2022})}\BibitemShut {NoStop}%
\bibitem [{\citenamefont {Jeon}\ \emph {et~al.}(2022)\citenamefont {Jeon},
  \citenamefont {Kim}, \citenamefont {Yoon}, \citenamefont {Jeon},
  \citenamefont {Han}, \citenamefont {Cottet}, \citenamefont {Kontos},\ and\
  \citenamefont {Parkin}}]{jeon_zero-field_2022}%
  \BibitemOpen
  \bibfield  {author} {\bibinfo {author} {\bibfnamefont {K.-R.}\ \bibnamefont
  {Jeon}}, \bibinfo {author} {\bibfnamefont {J.-K.}\ \bibnamefont {Kim}},
  \bibinfo {author} {\bibfnamefont {J.}~\bibnamefont {Yoon}}, \bibinfo {author}
  {\bibfnamefont {J.-C.}\ \bibnamefont {Jeon}}, \bibinfo {author}
  {\bibfnamefont {H.}~\bibnamefont {Han}}, \bibinfo {author} {\bibfnamefont
  {A.}~\bibnamefont {Cottet}}, \bibinfo {author} {\bibfnamefont
  {T.}~\bibnamefont {Kontos}},\ and\ \bibinfo {author} {\bibfnamefont
  {S.~S.~P.}\ \bibnamefont {Parkin}},\ }\bibfield  {title} {\bibinfo {title}
  {Zero-field polarity-reversible {Josephson} supercurrent diodes enabled by a
  proximity-magnetized {Pt} barrier},\ }\href
  {https://doi.org/10.1038/s41563-022-01300-7} {\bibfield  {journal} {\bibinfo
  {journal} {Nature Materials}\ }\textbf {\bibinfo {volume} {21}},\ \bibinfo
  {pages} {1008} (\bibinfo {year} {2022})}\BibitemShut {NoStop}%
\bibitem [{\citenamefont {Pakizer}\ \emph {et~al.}(2021)\citenamefont
  {Pakizer}, \citenamefont {Scharf},\ and\ \citenamefont
  {Matos-Abiague}}]{Pakizer2021:PRR}%
  \BibitemOpen
  \bibfield  {author} {\bibinfo {author} {\bibfnamefont {J.~D.}\ \bibnamefont
  {Pakizer}}, \bibinfo {author} {\bibfnamefont {B.}~\bibnamefont {Scharf}},\
  and\ \bibinfo {author} {\bibfnamefont {A.}~\bibnamefont {Matos-Abiague}},\
  }\bibfield  {title} {\bibinfo {title} {Crystalline anisotropic topological
  superconductivity in planar josephson junctions},\ }\href
  {https://doi.org/10.1103/PhysRevResearch.3.013198} {\bibfield  {journal}
  {\bibinfo  {journal} {Phys. Rev. Res.}\ }\textbf {\bibinfo {volume} {3}},\
  \bibinfo {pages} {013198} (\bibinfo {year} {2021})}\BibitemShut {NoStop}%
\bibitem [{\citenamefont {Pekerten}\ \emph {et~al.}(2022)\citenamefont
  {Pekerten}, \citenamefont {Pakizer}, \citenamefont {Hawn},\ and\
  \citenamefont {Matos-Abiague}}]{Pekerten2022:PRB}%
  \BibitemOpen
  \bibfield  {author} {\bibinfo {author} {\bibfnamefont {B.}~\bibnamefont
  {Pekerten}}, \bibinfo {author} {\bibfnamefont {J.~D.}\ \bibnamefont
  {Pakizer}}, \bibinfo {author} {\bibfnamefont {B.}~\bibnamefont {Hawn}},\ and\
  \bibinfo {author} {\bibfnamefont {A.}~\bibnamefont {Matos-Abiague}},\
  }\bibfield  {title} {\bibinfo {title} {Anisotropic topological
  superconductivity in josephson junctions},\ }\href
  {https://doi.org/10.1103/PhysRevB.105.054504} {\bibfield  {journal} {\bibinfo
   {journal} {Phys. Rev. B}\ }\textbf {\bibinfo {volume} {105}},\ \bibinfo
  {pages} {054504} (\bibinfo {year} {2022})}\BibitemShut {NoStop}%
\bibitem [{\citenamefont {Shabani}\ \emph {et~al.}(2016)\citenamefont
  {Shabani}, \citenamefont {Kjaergaard}, \citenamefont {Suominen},
  \citenamefont {Kim}, \citenamefont {Nichele}, \citenamefont {Pakrouski},
  \citenamefont {Stankevic}, \citenamefont {Lutchyn}, \citenamefont
  {Krogstrup}, \citenamefont {Feidenhans'l}, \citenamefont {Kraemer},
  \citenamefont {Nayak}, \citenamefont {Troyer}, \citenamefont {Marcus},\ and\
  \citenamefont {Palmstrøm}}]{shabani_two-dimensional_2016}%
  \BibitemOpen
  \bibfield  {author} {\bibinfo {author} {\bibfnamefont {J.}~\bibnamefont
  {Shabani}}, \bibinfo {author} {\bibfnamefont {M.}~\bibnamefont {Kjaergaard}},
  \bibinfo {author} {\bibfnamefont {H.~J.}\ \bibnamefont {Suominen}}, \bibinfo
  {author} {\bibfnamefont {Y.}~\bibnamefont {Kim}}, \bibinfo {author}
  {\bibfnamefont {F.}~\bibnamefont {Nichele}}, \bibinfo {author} {\bibfnamefont
  {K.}~\bibnamefont {Pakrouski}}, \bibinfo {author} {\bibfnamefont
  {T.}~\bibnamefont {Stankevic}}, \bibinfo {author} {\bibfnamefont {R.~M.}\
  \bibnamefont {Lutchyn}}, \bibinfo {author} {\bibfnamefont {P.}~\bibnamefont
  {Krogstrup}}, \bibinfo {author} {\bibfnamefont {R.}~\bibnamefont
  {Feidenhans'l}}, \bibinfo {author} {\bibfnamefont {S.}~\bibnamefont
  {Kraemer}}, \bibinfo {author} {\bibfnamefont {C.}~\bibnamefont {Nayak}},
  \bibinfo {author} {\bibfnamefont {M.}~\bibnamefont {Troyer}}, \bibinfo
  {author} {\bibfnamefont {C.~M.}\ \bibnamefont {Marcus}},\ and\ \bibinfo
  {author} {\bibfnamefont {C.~J.}\ \bibnamefont {Palmstrøm}},\ }\bibfield
  {title} {\bibinfo {title} {Two-dimensional epitaxial
  superconductor-semiconductor heterostructures: {A} platform for topological
  superconducting networks},\ }\href
  {https://doi.org/10.1103/PhysRevB.93.155402} {\bibfield  {journal} {\bibinfo
  {journal} {Physical Review B}\ }\textbf {\bibinfo {volume} {93}},\ \bibinfo
  {pages} {155402} (\bibinfo {year} {2016})}\BibitemShut {NoStop}%
\bibitem [{\citenamefont {Fornieri}\ \emph {et~al.}(2019)\citenamefont
  {Fornieri}, \citenamefont {Whiticar}, \citenamefont {Setiawan}, \citenamefont
  {Portoles}, \citenamefont {Drachmann}, \citenamefont {Keselman},
  \citenamefont {Gronin}, \citenamefont {Thomas}, \citenamefont {Wang},
  \citenamefont {Kallaher}, \citenamefont {Gardner}, \citenamefont {Berg},
  \citenamefont {Manfra}, \citenamefont {Stern}, \citenamefont {Marcus},\ and\
  \citenamefont {Nichele}}]{fornieri_evidence_2019}%
  \BibitemOpen
  \bibfield  {author} {\bibinfo {author} {\bibfnamefont {A.}~\bibnamefont
  {Fornieri}}, \bibinfo {author} {\bibfnamefont {A.~M.}\ \bibnamefont
  {Whiticar}}, \bibinfo {author} {\bibfnamefont {F.}~\bibnamefont {Setiawan}},
  \bibinfo {author} {\bibfnamefont {E.}~\bibnamefont {Portoles}}, \bibinfo
  {author} {\bibfnamefont {A.~C.~C.}\ \bibnamefont {Drachmann}}, \bibinfo
  {author} {\bibfnamefont {A.}~\bibnamefont {Keselman}}, \bibinfo {author}
  {\bibfnamefont {S.}~\bibnamefont {Gronin}}, \bibinfo {author} {\bibfnamefont
  {C.}~\bibnamefont {Thomas}}, \bibinfo {author} {\bibfnamefont
  {T.}~\bibnamefont {Wang}}, \bibinfo {author} {\bibfnamefont {R.}~\bibnamefont
  {Kallaher}}, \bibinfo {author} {\bibfnamefont {G.~C.}\ \bibnamefont
  {Gardner}}, \bibinfo {author} {\bibfnamefont {E.}~\bibnamefont {Berg}},
  \bibinfo {author} {\bibfnamefont {M.~J.}\ \bibnamefont {Manfra}}, \bibinfo
  {author} {\bibfnamefont {A.}~\bibnamefont {Stern}}, \bibinfo {author}
  {\bibfnamefont {C.~M.}\ \bibnamefont {Marcus}},\ and\ \bibinfo {author}
  {\bibfnamefont {F.}~\bibnamefont {Nichele}},\ }\bibfield  {title} {\bibinfo
  {title} {Evidence of topological superconductivity in planar {Josephson}
  junctions},\ }\href {https://doi.org/10.1038/s41586-019-1068-8} {\bibfield
  {journal} {\bibinfo  {journal} {Nature}\ }\textbf {\bibinfo {volume} {569}},\
  \bibinfo {pages} {89} (\bibinfo {year} {2019})}\BibitemShut {NoStop}%
\bibitem [{\citenamefont {Dartiailh}\ \emph
  {et~al.}(2021{\natexlab{a}})\citenamefont {Dartiailh}, \citenamefont {Mayer},
  \citenamefont {Yuan}, \citenamefont {Wickramasinghe}, \citenamefont
  {Matos-Abiague}, \citenamefont {Žutić},\ and\ \citenamefont
  {Shabani}}]{dartiailh_phase_2021}%
  \BibitemOpen
  \bibfield  {author} {\bibinfo {author} {\bibfnamefont {M.~C.}\ \bibnamefont
  {Dartiailh}}, \bibinfo {author} {\bibfnamefont {W.}~\bibnamefont {Mayer}},
  \bibinfo {author} {\bibfnamefont {J.}~\bibnamefont {Yuan}}, \bibinfo {author}
  {\bibfnamefont {K.~S.}\ \bibnamefont {Wickramasinghe}}, \bibinfo {author}
  {\bibfnamefont {A.}~\bibnamefont {Matos-Abiague}}, \bibinfo {author}
  {\bibfnamefont {I.}~\bibnamefont {Žutić}},\ and\ \bibinfo {author}
  {\bibfnamefont {J.}~\bibnamefont {Shabani}},\ }\bibfield  {title} {\bibinfo
  {title} {Phase {Signature} of {Topological} {Transition} in {Josephson}
  {Junctions}},\ }\href {https://doi.org/10.1103/PhysRevLett.126.036802}
  {\bibfield  {journal} {\bibinfo  {journal} {Physical Review Letters}\
  }\textbf {\bibinfo {volume} {126}},\ \bibinfo {pages} {036802} (\bibinfo
  {year} {2021}{\natexlab{a}})}\BibitemShut {NoStop}%
\bibitem [{\citenamefont {Banerjee}\ \emph {et~al.}(2023)\citenamefont
  {Banerjee}, \citenamefont {Geier}, \citenamefont {Rahman}, \citenamefont
  {Thomas}, \citenamefont {Wang}, \citenamefont {Manfra}, \citenamefont
  {Flensberg},\ and\ \citenamefont {Marcus}}]{Banerjee2023:arxiv}%
  \BibitemOpen
  \bibfield  {author} {\bibinfo {author} {\bibfnamefont {A.}~\bibnamefont
  {Banerjee}}, \bibinfo {author} {\bibfnamefont {M.}~\bibnamefont {Geier}},
  \bibinfo {author} {\bibfnamefont {M.~A.}\ \bibnamefont {Rahman}}, \bibinfo
  {author} {\bibfnamefont {C.}~\bibnamefont {Thomas}}, \bibinfo {author}
  {\bibfnamefont {T.}~\bibnamefont {Wang}}, \bibinfo {author} {\bibfnamefont
  {M.~J.}\ \bibnamefont {Manfra}}, \bibinfo {author} {\bibfnamefont
  {K.}~\bibnamefont {Flensberg}},\ and\ \bibinfo {author} {\bibfnamefont
  {C.~M.}\ \bibnamefont {Marcus}},\ }\bibfield  {title} {\bibinfo {title}
  {Phase {Asymmetry} of {Andreev} {Spectra} from {Cooper}-{Pair} {Momentum}},\
  }\href {https://doi.org/10.1103/PhysRevLett.131.196301} {\bibfield  {journal}
  {\bibinfo  {journal} {Physical Review Letters}\ }\textbf {\bibinfo {volume}
  {131}},\ \bibinfo {pages} {196301} (\bibinfo {year} {2023})}\BibitemShut
  {NoStop}%
\bibitem [{\citenamefont {Wickramasinghe}\ \emph {et~al.}(2018)\citenamefont
  {Wickramasinghe}, \citenamefont {Mayer}, \citenamefont {Yuan}, \citenamefont
  {Nguyen}, \citenamefont {Jiao}, \citenamefont {Manucharyan},\ and\
  \citenamefont {Shabani}}]{wickramasinghe_transport_2018}%
  \BibitemOpen
  \bibfield  {author} {\bibinfo {author} {\bibfnamefont {K.~S.}\ \bibnamefont
  {Wickramasinghe}}, \bibinfo {author} {\bibfnamefont {W.}~\bibnamefont
  {Mayer}}, \bibinfo {author} {\bibfnamefont {J.}~\bibnamefont {Yuan}},
  \bibinfo {author} {\bibfnamefont {T.}~\bibnamefont {Nguyen}}, \bibinfo
  {author} {\bibfnamefont {L.}~\bibnamefont {Jiao}}, \bibinfo {author}
  {\bibfnamefont {V.}~\bibnamefont {Manucharyan}},\ and\ \bibinfo {author}
  {\bibfnamefont {J.}~\bibnamefont {Shabani}},\ }\bibfield  {title} {\bibinfo
  {title} {Transport properties of near surface {InAs} two-dimensional
  heterostructures},\ }\href {https://doi.org/10.1063/1.5050413} {\bibfield
  {journal} {\bibinfo  {journal} {Applied Physics Letters}\ }\textbf {\bibinfo
  {volume} {113}},\ \bibinfo {pages} {262104} (\bibinfo {year}
  {2018})}\BibitemShut {NoStop}%
\bibitem [{\citenamefont {Strickland}\ \emph {et~al.}(2022)\citenamefont
  {Strickland}, \citenamefont {Hatefipour}, \citenamefont {Langone},
  \citenamefont {Farzaneh},\ and\ \citenamefont
  {Shabani}}]{strickland_controlling_2022}%
  \BibitemOpen
  \bibfield  {author} {\bibinfo {author} {\bibfnamefont {W.~M.}\ \bibnamefont
  {Strickland}}, \bibinfo {author} {\bibfnamefont {M.}~\bibnamefont
  {Hatefipour}}, \bibinfo {author} {\bibfnamefont {D.}~\bibnamefont {Langone}},
  \bibinfo {author} {\bibfnamefont {S.~M.}\ \bibnamefont {Farzaneh}},\ and\
  \bibinfo {author} {\bibfnamefont {J.}~\bibnamefont {Shabani}},\ }\bibfield
  {title} {\bibinfo {title} {Controlling {Fermi} level pinning in near-surface
  {InAs} quantum wells},\ }\href {https://doi.org/10.1063/5.0101579} {\bibfield
   {journal} {\bibinfo  {journal} {Applied Physics Letters}\ }\textbf {\bibinfo
  {volume} {121}},\ \bibinfo {pages} {092104} (\bibinfo {year}
  {2022})}\BibitemShut {NoStop}%
\bibitem [{\citenamefont {Mayer}\ \emph {et~al.}(2019)\citenamefont {Mayer},
  \citenamefont {Yuan}, \citenamefont {Wickramasinghe}, \citenamefont {Nguyen},
  \citenamefont {Dartiailh},\ and\ \citenamefont
  {Shabani}}]{mayer_superconducting_2019}%
  \BibitemOpen
  \bibfield  {author} {\bibinfo {author} {\bibfnamefont {W.}~\bibnamefont
  {Mayer}}, \bibinfo {author} {\bibfnamefont {J.}~\bibnamefont {Yuan}},
  \bibinfo {author} {\bibfnamefont {K.~S.}\ \bibnamefont {Wickramasinghe}},
  \bibinfo {author} {\bibfnamefont {T.}~\bibnamefont {Nguyen}}, \bibinfo
  {author} {\bibfnamefont {M.~C.}\ \bibnamefont {Dartiailh}},\ and\ \bibinfo
  {author} {\bibfnamefont {J.}~\bibnamefont {Shabani}},\ }\bibfield  {title}
  {\bibinfo {title} {Superconducting proximity effect in epitaxial al-{InAs}
  heterostructures},\ }\href {https://doi.org/10.1063/1.5067363} {\bibfield
  {journal} {\bibinfo  {journal} {Applied Physics Letters}\ }\textbf {\bibinfo
  {volume} {114}},\ \bibinfo {pages} {103104} (\bibinfo {year}
  {2019})}\BibitemShut {NoStop}%
\bibitem [{\citenamefont {Mayer}\ \emph {et~al.}(2020)\citenamefont {Mayer},
  \citenamefont {Dartiailh}, \citenamefont {Yuan}, \citenamefont
  {Wickramasinghe}, \citenamefont {Rossi},\ and\ \citenamefont
  {Shabani}}]{mayer_gate_2020}%
  \BibitemOpen
  \bibfield  {author} {\bibinfo {author} {\bibfnamefont {W.}~\bibnamefont
  {Mayer}}, \bibinfo {author} {\bibfnamefont {M.~C.}\ \bibnamefont
  {Dartiailh}}, \bibinfo {author} {\bibfnamefont {J.}~\bibnamefont {Yuan}},
  \bibinfo {author} {\bibfnamefont {K.~S.}\ \bibnamefont {Wickramasinghe}},
  \bibinfo {author} {\bibfnamefont {E.}~\bibnamefont {Rossi}},\ and\ \bibinfo
  {author} {\bibfnamefont {J.}~\bibnamefont {Shabani}},\ }\bibfield  {title}
  {\bibinfo {title} {Gate controlled anomalous phase shift in al/inas josephson
  junctions},\ }\href {https://doi.org/10.1038/s41467-019-14094-1} {\bibfield
  {journal} {\bibinfo  {journal} {Nature Communications}\ }\textbf {\bibinfo
  {volume} {11}},\ \bibinfo {pages} {212} (\bibinfo {year} {2020})}\BibitemShut
  {NoStop}%
\bibitem [{\citenamefont {Dartiailh}\ \emph
  {et~al.}(2021{\natexlab{b}})\citenamefont {Dartiailh}, \citenamefont
  {Cuozzo}, \citenamefont {Elfeky}, \citenamefont {Mayer}, \citenamefont
  {Yuan}, \citenamefont {Wickramasinghe}, \citenamefont {Rossi},\ and\
  \citenamefont {Shabani}}]{dartiailh_missing_2021}%
  \BibitemOpen
  \bibfield  {author} {\bibinfo {author} {\bibfnamefont {M.~C.}\ \bibnamefont
  {Dartiailh}}, \bibinfo {author} {\bibfnamefont {J.~J.}\ \bibnamefont
  {Cuozzo}}, \bibinfo {author} {\bibfnamefont {B.~H.}\ \bibnamefont {Elfeky}},
  \bibinfo {author} {\bibfnamefont {W.}~\bibnamefont {Mayer}}, \bibinfo
  {author} {\bibfnamefont {J.}~\bibnamefont {Yuan}}, \bibinfo {author}
  {\bibfnamefont {K.~S.}\ \bibnamefont {Wickramasinghe}}, \bibinfo {author}
  {\bibfnamefont {E.}~\bibnamefont {Rossi}},\ and\ \bibinfo {author}
  {\bibfnamefont {J.}~\bibnamefont {Shabani}},\ }\bibfield  {title} {\bibinfo
  {title} {Missing shapiro steps in topologically trivial josephson junction on
  {InAs} quantum well},\ }\href {https://doi.org/10.1038/s41467-020-20382-y}
  {\bibfield  {journal} {\bibinfo  {journal} {Nature Communications}\ }\textbf
  {\bibinfo {volume} {12}},\ \bibinfo {pages} {78} (\bibinfo {year}
  {2021}{\natexlab{b}})}\BibitemShut {NoStop}%
\bibitem [{\citenamefont {Courtois}\ \emph {et~al.}(2008)\citenamefont
  {Courtois}, \citenamefont {Meschke}, \citenamefont {Peltonen},\ and\
  \citenamefont {Pekola}}]{courtois_origin_2008}%
  \BibitemOpen
  \bibfield  {author} {\bibinfo {author} {\bibfnamefont {H.}~\bibnamefont
  {Courtois}}, \bibinfo {author} {\bibfnamefont {M.}~\bibnamefont {Meschke}},
  \bibinfo {author} {\bibfnamefont {J.~T.}\ \bibnamefont {Peltonen}},\ and\
  \bibinfo {author} {\bibfnamefont {J.~P.}\ \bibnamefont {Pekola}},\ }\bibfield
   {title} {\bibinfo {title} {Origin of {Hysteresis} in a {Proximity}
  {Josephson} {Junction}},\ }\href
  {https://doi.org/10.1103/PhysRevLett.101.067002} {\bibfield  {journal}
  {\bibinfo  {journal} {Physical Review Letters}\ }\textbf {\bibinfo {volume}
  {101}},\ \bibinfo {pages} {067002} (\bibinfo {year} {2008})}\BibitemShut
  {NoStop}%
\bibitem [{\citenamefont {Farzaneh}\ \emph {et~al.}(2024)\citenamefont
  {Farzaneh}, \citenamefont {Hatefipour}, \citenamefont {Schiela},
  \citenamefont {Lotfizadeh}, \citenamefont {Yu}, \citenamefont {Elfeky},
  \citenamefont {Strickland}, \citenamefont {Matos-Abiague},\ and\
  \citenamefont {Shabani}}]{farzaneh_magneto_anisotropic_2022}%
  \BibitemOpen
  \bibfield  {author} {\bibinfo {author} {\bibfnamefont {S.~M.}\ \bibnamefont
  {Farzaneh}}, \bibinfo {author} {\bibfnamefont {M.}~\bibnamefont
  {Hatefipour}}, \bibinfo {author} {\bibfnamefont {W.~F.}\ \bibnamefont
  {Schiela}}, \bibinfo {author} {\bibfnamefont {N.}~\bibnamefont {Lotfizadeh}},
  \bibinfo {author} {\bibfnamefont {P.}~\bibnamefont {Yu}}, \bibinfo {author}
  {\bibfnamefont {B.~H.}\ \bibnamefont {Elfeky}}, \bibinfo {author}
  {\bibfnamefont {W.~M.}\ \bibnamefont {Strickland}}, \bibinfo {author}
  {\bibfnamefont {A.}~\bibnamefont {Matos-Abiague}},\ and\ \bibinfo {author}
  {\bibfnamefont {J.}~\bibnamefont {Shabani}},\ }\bibfield  {title} {\bibinfo
  {title} {Observing magnetoanisotropic weak antilocalization in near-surface
  quantum wells},\ }\href {https://doi.org/10.1103/PhysRevResearch.6.013039}
  {\bibfield  {journal} {\bibinfo  {journal} {Physical Review Research}\
  }\textbf {\bibinfo {volume} {6}},\ \bibinfo {pages} {013039} (\bibinfo {year}
  {2024})},\ \Eprint {https://arxiv.org/abs/2208.06050} {2208.06050}
  \BibitemShut {NoStop}%
\bibitem [{\citenamefont {Costa}\ \emph
  {et~al.}(2023{\natexlab{b}})\citenamefont {Costa}, \citenamefont {Fabian},\
  and\ \citenamefont {Kochan}}]{Costa2023:arXiv}%
  \BibitemOpen
  \bibfield  {author} {\bibinfo {author} {\bibfnamefont {A.}~\bibnamefont
  {Costa}}, \bibinfo {author} {\bibfnamefont {J.}~\bibnamefont {Fabian}},\ and\
  \bibinfo {author} {\bibfnamefont {D.}~\bibnamefont {Kochan}},\ }\bibfield
  {title} {\bibinfo {title} {Microscopic study of the {Josephson} supercurrent
  diode effect in {Josephson} junctions based on two-dimensional electron
  gas},\ }\href {https://doi.org/10.1103/PhysRevB.108.054522} {\bibfield
  {journal} {\bibinfo  {journal} {Physical Review B}\ }\textbf {\bibinfo
  {volume} {108}},\ \bibinfo {pages} {054522} (\bibinfo {year}
  {2023}{\natexlab{b}})}\BibitemShut {NoStop}%
\bibitem [{\citenamefont {Lo}\ \emph {et~al.}(2013)\citenamefont {Lo},
  \citenamefont {Chen}, \citenamefont {Lin}, \citenamefont {Wu}, \citenamefont
  {Lin}, \citenamefont {Yeh}, \citenamefont {Chen},\ and\ \citenamefont
  {Liang}}]{Lo2013:SR}%
  \BibitemOpen
  \bibfield  {author} {\bibinfo {author} {\bibfnamefont {S.-T.}\ \bibnamefont
  {Lo}}, \bibinfo {author} {\bibfnamefont {K.~Y.}\ \bibnamefont {Chen}},
  \bibinfo {author} {\bibfnamefont {S.}~\bibnamefont {Lin}}, \bibinfo {author}
  {\bibfnamefont {J.-Y.}\ \bibnamefont {Wu}}, \bibinfo {author} {\bibfnamefont
  {T.~L.}\ \bibnamefont {Lin}}, \bibinfo {author} {\bibfnamefont {M.~R.}\
  \bibnamefont {Yeh}}, \bibinfo {author} {\bibfnamefont {T.-M.}\ \bibnamefont
  {Chen}},\ and\ \bibinfo {author} {\bibfnamefont {C.-T.}\ \bibnamefont
  {Liang}},\ }\bibfield  {title} {\bibinfo {title} {Controllable disorder in a
  hybrid nanoelectronic system: Realization of a superconducting diode},\
  }\href {https://doi.org/https://doi.org/10.1038/srep02274} {\bibfield
  {journal} {\bibinfo  {journal} {Sci. Rep}\ }\textbf {\bibinfo {volume} {3}},\
  \bibinfo {pages} {2274} (\bibinfo {year} {2013})}\BibitemShut {NoStop}%
\bibitem [{\citenamefont {Lustikova}\ \emph {et~al.}(2018)\citenamefont
  {Lustikova}, \citenamefont {Shiomi}, \citenamefont {Yokoi}, \citenamefont
  {Kabeya}, \citenamefont {Kimura}, \citenamefont {Ienaga}, \citenamefont
  {Kaneko}, \citenamefont {Okuma}, \citenamefont {Takahashi},\ and\
  \citenamefont {Saitoh}}]{Lustikova2018:NC}%
  \BibitemOpen
  \bibfield  {author} {\bibinfo {author} {\bibfnamefont {J.}~\bibnamefont
  {Lustikova}}, \bibinfo {author} {\bibfnamefont {Y.}~\bibnamefont {Shiomi}},
  \bibinfo {author} {\bibfnamefont {N.}~\bibnamefont {Yokoi}}, \bibinfo
  {author} {\bibfnamefont {N.}~\bibnamefont {Kabeya}}, \bibinfo {author}
  {\bibfnamefont {N.}~\bibnamefont {Kimura}}, \bibinfo {author} {\bibfnamefont
  {K.}~\bibnamefont {Ienaga}}, \bibinfo {author} {\bibfnamefont {S.-i.}\
  \bibnamefont {Kaneko}}, \bibinfo {author} {\bibfnamefont {S.}~\bibnamefont
  {Okuma}}, \bibinfo {author} {\bibfnamefont {S.}~\bibnamefont {Takahashi}},\
  and\ \bibinfo {author} {\bibfnamefont {E.}~\bibnamefont {Saitoh}},\
  }\bibfield  {title} {\bibinfo {title} {Vortex rectenna powered by
  environmental fluctuations},\ }\href@noop {} {\bibfield  {journal} {\bibinfo
  {journal} {Nat. Commun.}\ }\textbf {\bibinfo {volume} {9}},\ \bibinfo {pages}
  {4922} (\bibinfo {year} {2018})}\BibitemShut {NoStop}%
\bibitem [{\citenamefont {Zhang}\ \emph {et~al.}(2020)\citenamefont {Zhang},
  \citenamefont {Xu}, \citenamefont {Zou}, \citenamefont {Ai}, \citenamefont
  {Dong}, \citenamefont {Huang}, \citenamefont {Leng}, \citenamefont {Liu},
  \citenamefont {Zhang}, \citenamefont {Jia}, \citenamefont {Peng},
  \citenamefont {Zhao}, \citenamefont {Yang}, \citenamefont {Li}, \citenamefont
  {Guo}, \citenamefont {J.~Haigh}, \citenamefont {Nagaosa}, \citenamefont
  {Shen},\ and\ \citenamefont {Xiu}}]{Zhang2020:NC}%
  \BibitemOpen
  \bibfield  {author} {\bibinfo {author} {\bibfnamefont {E.}~\bibnamefont
  {Zhang}}, \bibinfo {author} {\bibfnamefont {X.}~\bibnamefont {Xu}}, \bibinfo
  {author} {\bibfnamefont {Y.-C.}\ \bibnamefont {Zou}}, \bibinfo {author}
  {\bibfnamefont {L.}~\bibnamefont {Ai}}, \bibinfo {author} {\bibfnamefont
  {X.}~\bibnamefont {Dong}}, \bibinfo {author} {\bibfnamefont {C.}~\bibnamefont
  {Huang}}, \bibinfo {author} {\bibfnamefont {P.}~\bibnamefont {Leng}},
  \bibinfo {author} {\bibfnamefont {S.}~\bibnamefont {Liu}}, \bibinfo {author}
  {\bibfnamefont {Y.}~\bibnamefont {Zhang}}, \bibinfo {author} {\bibfnamefont
  {Z.}~\bibnamefont {Jia}}, \bibinfo {author} {\bibfnamefont {X.}~\bibnamefont
  {Peng}}, \bibinfo {author} {\bibfnamefont {M.}~\bibnamefont {Zhao}}, \bibinfo
  {author} {\bibfnamefont {Y.}~\bibnamefont {Yang}}, \bibinfo {author}
  {\bibfnamefont {Z.}~\bibnamefont {Li}}, \bibinfo {author} {\bibfnamefont
  {H.}~\bibnamefont {Guo}}, \bibinfo {author} {\bibfnamefont {S.}~\bibnamefont
  {J.~Haigh}}, \bibinfo {author} {\bibfnamefont {N.}~\bibnamefont {Nagaosa}},
  \bibinfo {author} {\bibfnamefont {J.}~\bibnamefont {Shen}},\ and\ \bibinfo
  {author} {\bibfnamefont {F.}~\bibnamefont {Xiu}},\ }\bibfield  {title}
  {\bibinfo {title} {Nonreciprocal superconducting nbse2 antenna},\ }\href@noop
  {} {\bibfield  {journal} {\bibinfo  {journal} {Nat. Commun.}\ }\textbf
  {\bibinfo {volume} {11}},\ \bibinfo {pages} {5634} (\bibinfo {year}
  {2020})}\BibitemShut {NoStop}%
\bibitem [{\citenamefont {Masuko}\ \emph {et~al.}(2022)\citenamefont {Masuko},
  \citenamefont {Kawamura}, \citenamefont {Yoshimi}, \citenamefont {Hirayama},
  \citenamefont {Ikeda}, \citenamefont {Watanabe}, \citenamefont {He},
  \citenamefont {Maryenko}, \citenamefont {Tsukazaki}, \citenamefont
  {Takahashi}, \citenamefont {Kawasaki}, \citenamefont {Nagaosa},\ and\
  \citenamefont {Tokura}}]{Masuko2022:npjQM}%
  \BibitemOpen
  \bibfield  {author} {\bibinfo {author} {\bibfnamefont {M.}~\bibnamefont
  {Masuko}}, \bibinfo {author} {\bibfnamefont {M.}~\bibnamefont {Kawamura}},
  \bibinfo {author} {\bibfnamefont {R.}~\bibnamefont {Yoshimi}}, \bibinfo
  {author} {\bibfnamefont {M.}~\bibnamefont {Hirayama}}, \bibinfo {author}
  {\bibfnamefont {Y.}~\bibnamefont {Ikeda}}, \bibinfo {author} {\bibfnamefont
  {R.}~\bibnamefont {Watanabe}}, \bibinfo {author} {\bibfnamefont {J.~J.}\
  \bibnamefont {He}}, \bibinfo {author} {\bibfnamefont {D.}~\bibnamefont
  {Maryenko}}, \bibinfo {author} {\bibfnamefont {A.}~\bibnamefont {Tsukazaki}},
  \bibinfo {author} {\bibfnamefont {K.~S.}\ \bibnamefont {Takahashi}}, \bibinfo
  {author} {\bibfnamefont {M.}~\bibnamefont {Kawasaki}}, \bibinfo {author}
  {\bibfnamefont {N.}~\bibnamefont {Nagaosa}},\ and\ \bibinfo {author}
  {\bibfnamefont {Y.}~\bibnamefont {Tokura}},\ }\bibfield  {title} {\bibinfo
  {title} {Nonreciprocal charge transport in topological superconductor
  candidate bi2te3/pdte2 heterostructure},\ }\href@noop {} {\bibfield
  {journal} {\bibinfo  {journal} {npj Quantum Mater.}\ }\textbf {\bibinfo
  {volume} {7}},\ \bibinfo {pages} {104} (\bibinfo {year} {2022})}\BibitemShut
  {NoStop}%
\bibitem [{\citenamefont {Eschrig}(2015)}]{Eschrig2015:RPP}%
  \BibitemOpen
  \bibfield  {author} {\bibinfo {author} {\bibfnamefont {M.}~\bibnamefont
  {Eschrig}},\ }\bibfield  {title} {\bibinfo {title} {Spin-polarized
  supercurrents for spintronics: a review of current progress},\ }\href
  {https://doi.org/10.1088/0034-4885/78/10/104501} {\bibfield  {journal}
  {\bibinfo  {journal} {Rep. Prog. Phys.}\ }\textbf {\bibinfo {volume} {78}},\
  \bibinfo {pages} {104501} (\bibinfo {year} {2015})}\BibitemShut {NoStop}%
\bibitem [{\citenamefont {Pientka}\ \emph {et~al.}(2017)\citenamefont
  {Pientka}, \citenamefont {Keselman}, \citenamefont {Berg}, \citenamefont
  {Yacoby}, \citenamefont {Stern},\ and\ \citenamefont
  {Halperin}}]{pientka_topological_2017}%
  \BibitemOpen
  \bibfield  {author} {\bibinfo {author} {\bibfnamefont {F.}~\bibnamefont
  {Pientka}}, \bibinfo {author} {\bibfnamefont {A.}~\bibnamefont {Keselman}},
  \bibinfo {author} {\bibfnamefont {E.}~\bibnamefont {Berg}}, \bibinfo {author}
  {\bibfnamefont {A.}~\bibnamefont {Yacoby}}, \bibinfo {author} {\bibfnamefont
  {A.}~\bibnamefont {Stern}},\ and\ \bibinfo {author} {\bibfnamefont {B.~I.}\
  \bibnamefont {Halperin}},\ }\bibfield  {title} {\bibinfo {title} {Topological
  {Superconductivity} in a {Planar} {Josephson} {Junction}},\ }\href
  {https://doi.org/10.1103/PhysRevX.7.021032} {\bibfield  {journal} {\bibinfo
  {journal} {Phys. Rev. X}\ }\textbf {\bibinfo {volume} {7}},\ \bibinfo {pages}
  {021032} (\bibinfo {year} {2017})}\BibitemShut {NoStop}%
\bibitem [{\citenamefont {Setiawan}\ \emph {et~al.}(2019)\citenamefont
  {Setiawan}, \citenamefont {Stern},\ and\ \citenamefont
  {Berg}}]{Setiawan2019:PRB}%
  \BibitemOpen
  \bibfield  {author} {\bibinfo {author} {\bibfnamefont {F.}~\bibnamefont
  {Setiawan}}, \bibinfo {author} {\bibfnamefont {A.}~\bibnamefont {Stern}},\
  and\ \bibinfo {author} {\bibfnamefont {E.}~\bibnamefont {Berg}},\ }\bibfield
  {title} {\bibinfo {title} {Topological superconductivity in planar josephson
  junctions: Narrowing down to the nanowire limit},\ }\href
  {https://doi.org/10.1103/PhysRevB.99.220506} {\bibfield  {journal} {\bibinfo
  {journal} {Phys. Rev. B}\ }\textbf {\bibinfo {volume} {99}},\ \bibinfo
  {pages} {220506(R)} (\bibinfo {year} {2019})}\BibitemShut {NoStop}%
\bibitem [{\citenamefont {Pakizer}\ and\ \citenamefont
  {Matos-Abiague}(2021)}]{Pakizer2021:PRB}%
  \BibitemOpen
  \bibfield  {author} {\bibinfo {author} {\bibfnamefont {J.~D.}\ \bibnamefont
  {Pakizer}}\ and\ \bibinfo {author} {\bibfnamefont {A.}~\bibnamefont
  {Matos-Abiague}},\ }\bibfield  {title} {\bibinfo {title} {Signatures of
  topological transitions in the spin susceptibility of josephson junctions},\
  }\href {https://doi.org/10.1103/PhysRevB.104.L100506} {\bibfield  {journal}
  {\bibinfo  {journal} {Phys. Rev. B}\ }\textbf {\bibinfo {volume} {104}},\
  \bibinfo {pages} {L100506} (\bibinfo {year} {2021})}\BibitemShut {NoStop}%
\bibitem [{\citenamefont {Schiela}\ \emph {et~al.}(2024)\citenamefont
  {Schiela}, \citenamefont {Lotfizadeh}, \citenamefont {Pekerten},
  \citenamefont {Yu}, \citenamefont {Elfeky}, \citenamefont {Strickland},
  \citenamefont {Matos-Abiague},\ and\ \citenamefont {Shabani}}]{datarepo}%
  \BibitemOpen
  \bibfield  {author} {\bibinfo {author} {\bibfnamefont {W.~F.}\ \bibnamefont
  {Schiela}}, \bibinfo {author} {\bibfnamefont {N.}~\bibnamefont {Lotfizadeh}},
  \bibinfo {author} {\bibfnamefont {B.}~\bibnamefont {Pekerten}}, \bibinfo
  {author} {\bibfnamefont {P.}~\bibnamefont {Yu}}, \bibinfo {author}
  {\bibfnamefont {B.~H.}\ \bibnamefont {Elfeky}}, \bibinfo {author}
  {\bibfnamefont {W.~M.}\ \bibnamefont {Strickland}}, \bibinfo {author}
  {\bibfnamefont {A.}~\bibnamefont {Matos-Abiague}},\ and\ \bibinfo {author}
  {\bibfnamefont {J.}~\bibnamefont {Shabani}},\ }\bibfield  {title} {\bibinfo
  {title} {{Data from ``Superconducting Diode Effect Sign Change in Epitaxial
  Al-InAs Josepshon Junctions''}},\ }\href
  {https://doi.org/10.5281/zenodo.10810819} {10.5281/zenodo.10810819} (\bibinfo
  {year} {2024})\BibitemShut {NoStop}%
\end{thebibliography}%

\section{Acknowledgments}

This work is supported in part by DARPA Topological Excitations in Electronics (TEE) program under grant No. DP18AP900007, the U.S. Office of Naval Research (ONR) through Grants No. N000142112450 and MURI No. N000142212764. B.P and A.M.A. acknowledge support from ONR Grant No. N000141712793.

\section{Author contributions}
W.~M.~S.\ grew the material heterostructure.
N.~L.\ and P.~Y.\ fabricated the devices.
N.~L., W.~F.~S., and B.H.E\ performed the measurements.
N.~L.\ and W.~F.~S.\ performed the analysis.
B.~P.\ and A.~M.\ performed the analytical calculations and numerical simulations.
N.~L., W.~F.~S., A.~M., and B.~P.\ wrote the manuscript.
J.~S.\ conceived the experiment.

\section{Competing interests}
The authors declare no competing interests.

\end{document}